\documentclass[12pt]{article}
\usepackage{amsmath}
\usepackage{graphicx}%
\usepackage{amsfonts}%
\usepackage{amssymb}
\usepackage{xcolor}
\usepackage{placeins}
\usepackage{natbib}
\definecolor{myblue}{HTML}{377EB8}

\usepackage{xr}
\makeatletter
\newcommand*{\addFileDependency}[1]{
  \typeout{(#1)}
  \@addtofilelist{#1}
  \IfFileExists{#1}{}{\typeout{No file #1.}}
}
\makeatother

\newcommand*{\myexternaldocument}[1]{%
    \externaldocument{#1}%
    \addFileDependency{#1.tex}%
    \addFileDependency{#1.aux}%
}
\myexternaldocument{supp3}

\def\spacingset#1{\renewcommand{\baselinestretch}%
{#1}\small\normalsize} \spacingset{1}

\graphicspath{{Figures/}}

\begin{document}

\title{\bf Distributed lag models to identify the cumulative effects of training and recovery in athletes using multivariate ordinal wellness data}

\author{}
\author{Erin M. Schliep\\ Department of Statistics, University of Missouri\\ and\\ Toryn L.J. Schafer\\ Department of Statistics, University of Missouri\\ and\\ Matthew Hawkey\\ Institute of Sport, Exercise and Active Living, Victoria University Melbourne}
\date{} 
\maketitle
\bigskip
\begin{abstract}
Subjective wellness data can provide important information on the well-being of athletes and be used to maximize player performance and detect and prevent against injury. Wellness data, which are often ordinal and multivariate, include metrics relating to the physical, mental, and emotional status of the athlete. Training and recovery can have significant short- and long-term effects on athlete wellness, and these effects can vary across individual. We develop a joint multivariate latent factor model for ordinal response data to investigate the effects of training and recovery on athlete wellness. We use a latent factor distributed lag model to capture the cumulative effects of training and recovery through time. Current efforts using subjective wellness data have averaged over these metrics to create a univariate summary of wellness, however this approach can mask important information in the data. Our multivariate model leverages each ordinal variable and can be used to identify the relative importance of each in monitoring athlete wellness. The model is applied to athlete daily wellness, training, and recovery data collected across two Major League Soccer seasons.
\end{abstract}

\noindent%

\spacingset{1.5} 

\section{Introduction}
\label{Sec:Intro}

The rapid increase in data collection technology in sports over the previous decade has led to an evolution of player monitoring with regard to player performance, assessment, and injury detection and prevention \citep{Bourdon2017, Akenhead2016,De2018}. 
Within both individual and team sports, objective and subjective player monitoring data are commonly being used to assess the acute and chronic effects of training and recovery in order to maximize the current and future performance of athletes \citep{Mujika2017, Saw2016, Thorpe2015, Thorpe2017, Buchheit2013, Meeusen2013}. 

Customary metrics of an athlete's training response include objective measures of performance, physiology, or biochemistry, such as heart rate, heart rate variability, blood pressure, or oxygen consumption \citep{Borresen2009}. 
Recently, there has been strong emphasis on also including subjective, or perceptual, measures of well-being in athlete assessment \citep{Akenhead2016, Tavares2018}. 
Subjective measures, which are often self-reported by the athletes using self-assessment surveys, include wellness profiles to quantify physical, mental, and emotion state, sleep quality, energy levels and fatigue, and measures of perceived effort during training. 
Subjective measures have been reported to be more sensitive and consistent than objective measures in capturing acute and chronic training loads \citep{Saw2016, Tavares2018}. 
In particular, \cite{Saw2016} found subjective well-being to negatively respond to acute increases in training load as well as to chronic training, whereas acute decreases in training load led to an increases in subjective well-being. 
Subjective rate of perceived effort has also been reported to provide a reasonable assessment of training load compared to objective heart-rate based methods \citep{Borresen2008}, however accuracy varied as a function of the amount of high- and low-intensity exercise.
\cite{Tavares2018} investigated the effects of training and recovery on rugby athletes based on an objective measure of fatigue using countermovement jump tests as well as perceptual muscle-group specific measures of soreness for six days prior to a rugby match and two days following. 
They found that perceptual muscle soreness tended to diminish after a recovery day while lower body muscle soreness remained above the baseline.
In addition, even though running load, as measured by total distance and high metabolic load distance, did not differ between the match and training sessions, the change in perceptual measures of muscle soreness indicated greater physical demands on the athletes during rugby matches. 
The highest soreness scores for all muscle groups were reported the morning after the match, and these scores remained high the subsequent morning.
Interestingly, no significant difference in countermovement jump scores was detected across days. 
Since no one metric is best at quantifying the effects of training on fitness and fatigue and predicting performance \citep{Bourdon2017}, subjective and objective measures are often used in conjunction to guide training programs to improve athlete performance.  

Self-reported wellness measures are comprised of responses to a set of survey questions regarding the athlete's mental, emotional, and physical well-being. 
Most commonly, these survey responses are recorded on ordinal (or Likert) scales.
Collectively, these ordinal response variables inform on the athlete's overall well-being. 
Previous studies looked at the average of a set of ordinal wellness response variables and treated the average as a continuous response variable in a multiple linear regression model \citep{Gallo2017}.
Whereas this approach can be used as an exploratory tool to identify important training variables (work load, duration of matches/games, recovery) on individual wellness, it has three major shortcomings. 
First and foremost, it throws away important information by reducing the multiple wellness metrics into one value. 
Second, it assumes that each ordinal response variable is equally important (i.e., assigns weight $1/J$ for each $j=1, \dots, J$ variable). 
Since there is likely variation in the sensitivity of some wellness variables in an athlete's response to training and recovery, failing to differentiate between these variables could mask indicators of potential poor performance or negative health outcomes. 
Lastly, by modeling the average of the wellness metrics, the model is unable to identify important variable-specific relationships between training and wellness.

With the increase in collection of self-reported wellness measures and their identified significance in monitoring player performance, we need more advanced statistical methods and models that leverage the information across all wellness metrics in order to obtain a better understanding of the subjective measures of training and wellness. These models could then be used to synthesize these data in order to guide training programs and player evaluations.  

There are many statistical challenges in modeling subjective wellness data.
First, the data are multivariate, with each variable representing a particular aspect of wellness (e.g., energy levels, mental state, etc.).
Second, the data are ordinal, requiring more advanced generalized linear models of which are not often included in customary statistical programming packages \citep[although see the R package \texttt{mvord}][]{mvord}.
Lastly, the subjective wellness data are individual-specific. 
The day-to-day variation in wellness scores reported by an individual will vary greatly not only between individuals but also across wellness variable for an individual. 
In addition, training and recovery can have varying short- and long-term effects on athlete wellness, of which are also known to vary across individuals.
As such, statistical modeling of individual wellness needs to be able account for the variation across individual as a response to the cumulative effects of training and recovery.

Generalized latent variable models, including multilevel models and structural equation models, have been proposed as a comprehensive approach for modeling multivariate ordinal data and for capturing complex dependencies both between variables and within variables across time and space \citep{Skrondal2004}.
To capture flexible, nonlinear relationships, \cite{Deyoreo2018} developed a Bayesian nonparametric approach for multivariate ordinal regression based on mixture modeling for the joint distribution of latent responses and covariates. 
\cite{Schliep2013} developed a multi-level spatially-dependent latent factor model to assess the biotic condition of wetlands across a river basin using five ordinal response metrics.
A modified approach by \cite{Cagnone2018} used a latent Markov model to model temporal dynamics in the latent factor for multivariate longitudinal ordinal data.
\cite{Cagnone2009} propose a latent variable model with both a common latent factor and auto-regressive random effect to capture dependencies between the variables dynamically through time.
Within the class of generalized linear multivariate mixed models,
\cite{Chaubert2008} proposed a dynamic multivariate ordinal probit model by placing an auto-regressive structure on the coefficients and threshold parameters of the probit regression model.
Multivariate ordinal data are also common in the educational testing literature where mixed effects models (or item response theory models) are used to compare ordinal responses across individual \citep{Lord2012}.
\cite{Liu2006} developed a multi-level item response theory regression model for longitudinal multivariate ordinal data to study the substance use behaviors over time in inner-city youth.

Drawing on this literature, we propose using the latent factor model approach to capture marginal dependence between the multivariate ordinal wellness variables. 
We extend the current approaches by modeling the latent factors using distributed lag models to allow for functional effects of training and recovery.
Distributed lag models (also known as dynamic regression models) stem from the time series literature \citep[see][Chapter 9]{Hyndman2018} and offer an approach for identifying the dynamics relating two time series \citep{Haugh1977}.
Distributed lag models can be written as regression models in which a series of lagged explanatory variables accounts for the temporal variability in the response process. 
The coefficients of these lagged variables can be used to infer short- and long-term effects of important explanatory variables on the response. 
Due to the dependent relationship between the lagged values of the process, constraints are often imposed on the coefficients to induce shrinkage but maintain interpretability. 
Often the constraints impose a smooth functional relationship between the explanatory variables and response. 
The smoothed coefficients can then be interpreted as a time series of effect sizes and provide insights to the significance of the explanatory variables at various lags.
In epidemiological research, distributed lag models have been used to capture the lag time between exposure and response. For example, \cite{Schwartz2000} studied air pollution exposure on adverse health outcomes in humans and found up to a 5 day lag in exposure effects.
\cite{Gasparrini2010} developed the family of distributed lag non-linear models to study non-linear relationships between exposure and response. 
The added flexibility of their model enables the shape and temporal lag of the relationship to be captured simultaneously.

In the ecological context, \cite{Ogle2015} and \cite{Itter2019} presented a special case of distributed lag models to capture so-called `ecological memory'. 
Like distributed lag models, ecological memory models assign a non-negative measure, or weight, to multiple previous time points to capture the possible short- and long-term effects of environmental variables (e.g., climate variables, such as precipitation or temperature) on various environmental processes (e.g., species occupancy).
These models are able to identify not only significant past events on the environmental process of interest but also the length of ``memory" these environmental process have with respect to the environmental variables. 
An important distinction between ecological memory models and distributed lag models is that ecological memory models assume the short- and long-term effects to be consistent. 
That is, with non-negative weights, the relationship between the explanatory variable and the response is the same at all lags governed by the sign of the coefficient.
In modeling the short- and long-term effects of training and recovery on athlete wellness, we note that this limitation may be overly restrictive. 

We propose a joint multivariate ordinal response latent factor distributed lag model to tackle the challenges outlined above. 
The joint model specification enables the borrowing of strength across athletes with regard to capturing the general trends in training response. 
However, athlete-specific model components allow for individualized effects and relationships between wellness measures, training, and recovery.
The latent factors capture the temporal variability in athlete wellness as a function of training and recovery. 
In addition, the multivariate model allows for the latent factors to be informed by each of the self-reported wellness metrics, which alleviates the pre-processing of computing the average and treating it as a univariate response.
We model the latent factors using distributed lag models in order to capture the cumulative effects of training load and recovery on athlete wellness.

The remainder of the paper is outlined as follows.
In Section \ref{Sec:Data} we describe the ordinal athlete wellness data, including the metrics of workload and recovery. 
Summaries of the data as well as exploratory data analysis is included.
The multivariate ordinal response latent factor distributed lag model is developed in Section \ref{Sec:Model}.
Details with regard to the model specification, model inference, and important identifiability constraints are included.
The model is applied to the athlete data in Section \ref{Sec:Results} and important results and inference are discussed.
We conclude with a summary and discussion of future work in Section \ref{Sec:Disc}.

\section{Athlete wellness, training, and recovery data}
\label{Sec:Data}

Daily wellness, training, and recovery data were obtained for 20 professional referees during the 2015 and 2016 seasons of Major League Soccer (MLS), which spanned from approximately February 1$^{st}$ to October 30$^{th}$ of each year. 
Each referee followed a training program established by the Professional Referees Organization and attended bi-weekly training camps.
Over the span of these two seasons, the number of matches officiated by each referee ranged from 3 to 44, with an average of 28 matches.

Upon waking up each morning, the referee (hereafter, ``athlete") was prompted on his smart phone to complete a wellness survey. The survey questions entailed assigning a value to each wellness variable (metric) on a scale from 1 to 10 where 1 is low/worst and 10 is high/best.
These wellness variables include \emph{energy}, \emph{tiredness}, \emph{motivation}, \emph{stress}, \emph{mood}, and \emph{appetite}.
Prior to the start of data collection, the referees were trained on the ordinal scoring for these variables such that a high value of all variables corresponds to being energized, fully rested, feeling highly motivated, with low stress, a positive attitude, and being hungry. 

Due to the limited number of observations in each of the 10 categories for most individuals and wellness variables, we transformed the raw ordinal data to a 5 category scale. 
The transformation we elected to use was individual specific, but uniform across metrics as we assumed each individual's ability to differentiate between ordinal values was consistent across variables.
For each individual, we combined the observed ordinal data and conducted $k$-means clustering using 5 clusters. 
The $k$-means clustering technique minimizes within group variability and identifies a center value for each cluster.
The midpoints between the ordered cluster centers 
were used as cut-points to assign ordinal values from 1 to 5. 
Each of the six ordinal wellness variables for the individual was transformed to the 5 category ordinal scale using the same set of cut-points.
We investigated various transformation (e.g., basic combining of classes 1-2, 3-4, etc, and re-scaling the data to (0,1) and then applying a threshold based on percentiles 0.2, 0.4, etc.) but found model inference in general to be robust to these choices.

\begin{figure}
\begin{center}
\includegraphics[scale=.48]{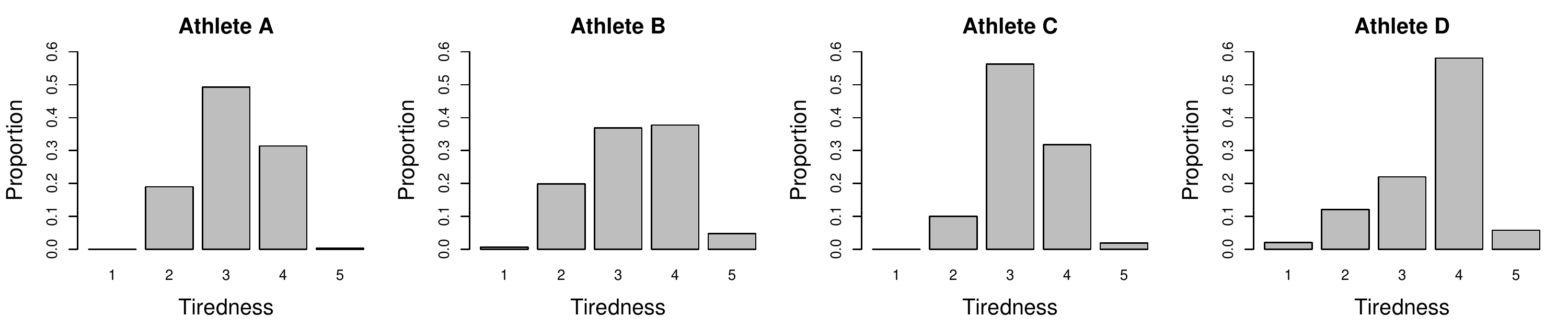}
\includegraphics[scale=.48]{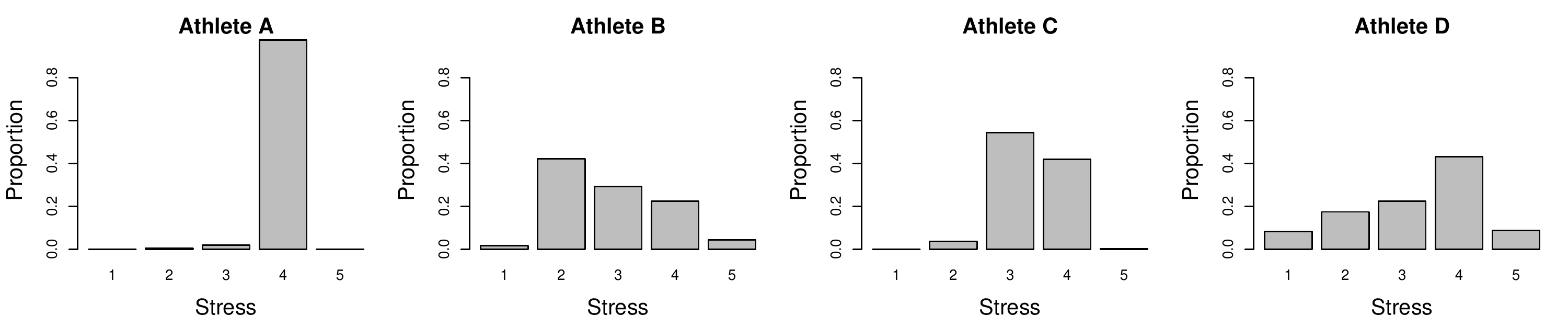}
\caption{Distributions of the ordinal wellness variables \emph{tiredness} (top) and \emph{stress} (bottom) for four athletes. \label{Fig:WM1}}
\end{center}
\end{figure}

Distributions of the 5-category ordinal data for the two metrics, \emph{tiredness} and \emph{stress}, are shown in Figure \ref{Fig:WM1} for four athletes denoted Athlete A, B, C, and D. 
The distributions of wellness scores vary quite drastically both across metrics for a given athlete as well as across athletes for a given metric. 
The distributions of the other wellness variables for these four athletes are included in the Figure \ref{Fig: WM2} of the Supplementary Material.

\begin{figure}
\begin{center}
\includegraphics[scale=.48]{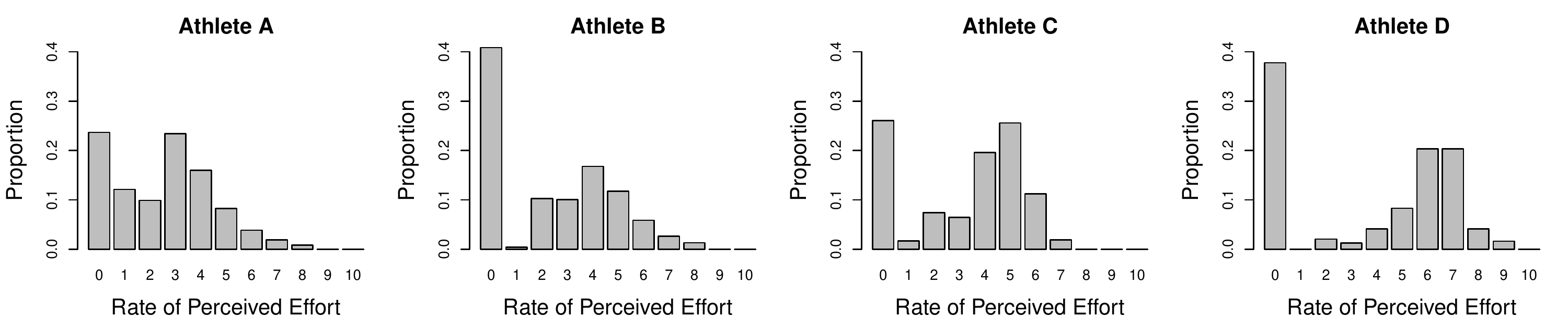}
\includegraphics[scale=.48]{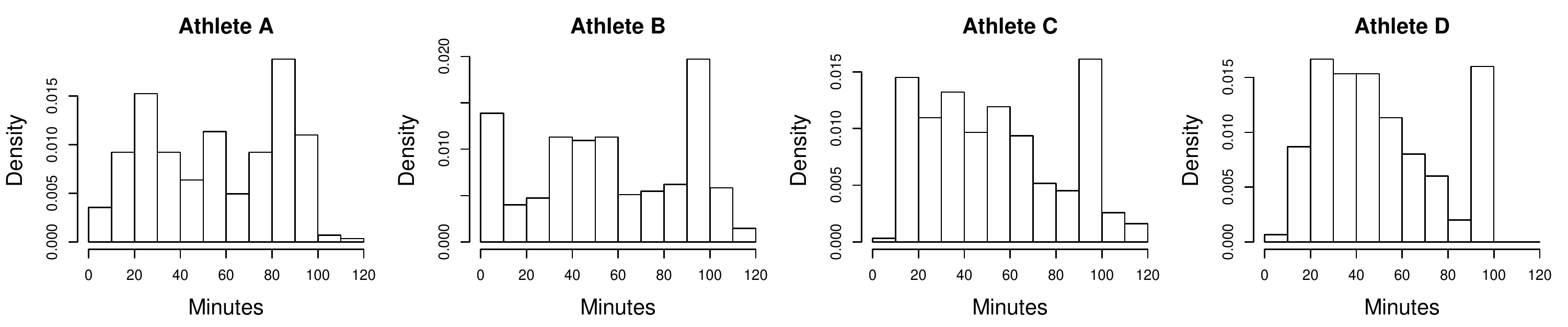}
\caption{Distributions of the rate of perceived effort (top) and workout duration (bottom) for four athletes in the data. Duration distributions include only workouts with a non-zero duration.\label{Fig:RPEDur}}
\end{center}
\end{figure}

In addition to the wellness variables, each athlete also reported information on training and recovery. 
These data consisted of the duration (hours) and rate of perceived effort (RPE; scored 0-10) of the previous days workout as well as sleep quantity (hours) and quality (scored 1-10) of the previous nights sleep. 
Distributions of RPE and workout duration are shown in Figure \ref{Fig:RPEDur} for the four athletes.
RPE has been found to be a valid method for quantifying training across a variety of types of exercise \citep{Foster2001, Haddad2017}.
The distribution of RPE varies across each individual both in terms of center and spread as well as with regard to the frequency of ``0" RPE days (i.e., rest days). 
For example, over 40\% of the days are reported as rest days for Athlete B. 
For non-rest days (those with RPE $>0$), Athlete D has a much higher average RPE compared to the other three athletes. 
The distribution of workout duration tends to be multi-modal for each athlete. 
In particular, we see spikes around 90 minutes, which is consistent with the length of MLS games. 
The distribution of shorter workouts for Athlete C tends to be more uniform between 10 and 70 minutes with fewer short workouts than the other athletes.

\begin{figure}
\begin{center}
\includegraphics[scale=.5]{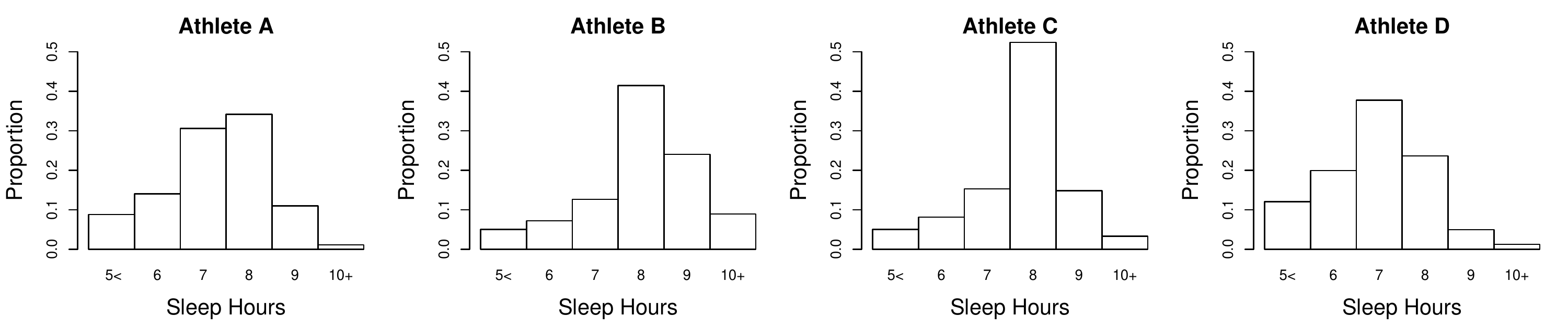}
\includegraphics[scale=.5]{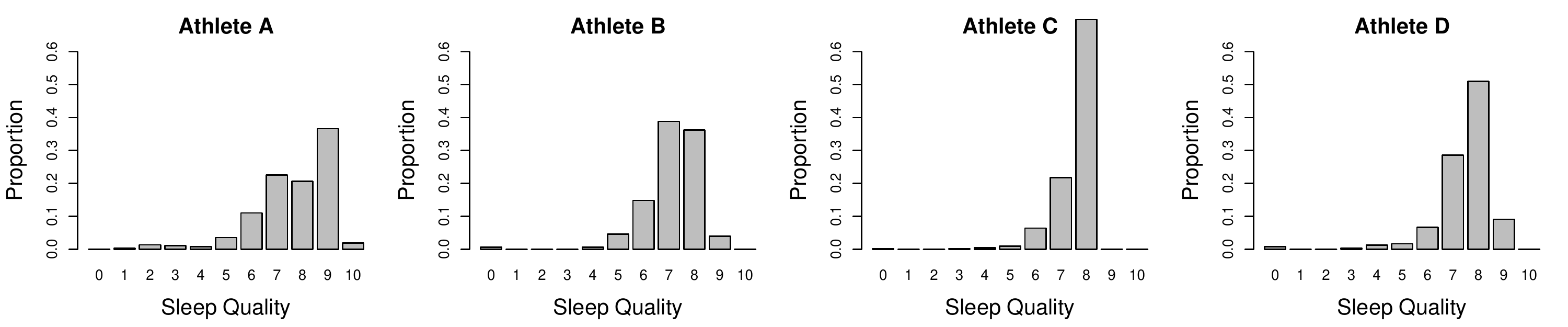}
\caption{Raw summaries of the number of hours slept (top) and quality of sleep (bottom) for four athletes in the data. \label{Fig:Sleep}}
\end{center}
\end{figure}

Similar summaries of the number of hours slept and the quality of sleep are shown in Figure \ref{Fig:Sleep}. 
In general, the average number of hours of sleep for an individual ranges between 6-8 hours, however the frequency of nights with 5 or less hours and 10 or more hours varies significantly across individuals. 
Sleep quality is most concentrated on values between 6 and 8 for each individual, with the distributions being skewed towards lower values.


We define two important measures of training and recovery to use in our modeling; namely, \emph{workload} and \emph{recovery}.
Workload, which is also commonly referred to as training load, is quantified as the product of the rate of perceived effort and the duration for the training period \citep{Foster1996, Foster2001, Brink2010}.
Therefore, a training session of moderate to high intensity and average to long duration will result in a large workload value. 
Recovery is defined using the reported quality and quantity of sleep.
We applied principle component analysis on the two sleep metrics for each individual and found that one loading vector captured between 64\% and 94\% of the variation. 
As a result, recovery for each athlete was defined using the first principle component, where large values correspond to overall good recovery (high quality and long duration sleep).

\section{Joint multivariate ordinal response latent factor distributed lag model}
\label{Sec:Model}
We model the ordinal wellness data using a multivariate ordinal response latent factor distributed lag model. 
We first describe the multivariate ordinal response model in Section \ref{Ord} and then offer two latent factor model specifications using distributed lag models in Section \ref{LFM}.
Identifiability constraints are discussed in Section \ref{sec:Ident}, and full prior specifications for Bayesian inference are given in Section \ref{sec:Inf1}. 
Section \ref{sec:Inf2} introduces important inference measures for addressing questions regarding athlete wellness, workload, and recovery.

\subsection{Multivariate ordinal response model}
\label{Ord}
Let $i = 1, \dots, n$ denote individual, $j=1, \dots, J$ denote wellness variable (which we refer to as \emph{metric}), and $t =1, \dots, T_i$ denote time (day). 
Then, define $Z_{ijt} \in \{1, 2, \dots, K_{ij}\}$ to be the ordinal value for individual $i$ and wellness metric $j$ on day $t$.
Without loss of generality, let $K_{ij}=5$ for each $i$ and $j$ such that each wellness metric is ordinal taking integer values $1, \dots, 5$ for each individual.

We model the ordinal response variables using a cumulative probit regression model.
We utilize the efficient parameterization of \cite{Albert1993}, and define the latent metric parameter $\widetilde{Z}_{ijt}$ such that
\begin{equation}
Z_{ijt} = \begin{cases} 
1 ~~~& -\infty < \widetilde{Z}_{ijt} \leq \theta^{(1)}_{ij}\\
2 ~~~&  \theta^{(1)}_{ij} < \widetilde{Z}_{ijt} \leq \theta^{(2)}_{ij}\\
3 ~~~&  \theta^{(2)}_{ij} < \widetilde{Z}_{ijt} \leq \theta^{(3)}_{ij}\\
4 ~~~&  \theta^{(3)}_{ij} < \widetilde{Z}_{ijt} \leq \theta^{(4)}_{ij}\\
5 ~~~&  \theta^{(4)}_{ij} < \widetilde{Z}_{ijt} < \infty. \\
\end{cases}
\end{equation}
Here, $\theta^{(k-1)}_{ij}$ and $\theta^{(k)}_{ij}$ denote the lower and upper thresholds of ordinal value $k$, for individual $i$ and wellness metric $j$, where $\theta^{(k-1)}_{ij} < \theta^{(k)}_{ij}$.
Under the general probit regression specification, 
\begin{equation}
\widetilde{Z}_{ijt} = \mu_{ijt} + \epsilon_{ijt}
\label{eq:Ztilde}
\end{equation} 
where $\epsilon_{ijt} \sim N(0,\sigma^2_{ij})$.
In the Bayesian framework with inference obtained using Markov chain Monte Carlo, this parameterization enables efficient Gibbs updates of the model parameters, $\widetilde{Z}_{ijt}$, $\mu_{ijt}$, and $\sigma^2_{ij}$, for all $i$, $j$, and $t$ \citep{Albert1993}. 
Posterior samples of the threshold parameters, $\theta^{(k)}_{ij}$, require a Metropolis step.
More details with regard to the sampling algorithm are given in Section \ref{sec:Ident}.

\subsection{Latent factor models}
\label{LFM}
In modeling $\mu_{ijt}$, we propose both a univariate and multivariate latent factor model specification to generate important, distinct inferential measures.
We begin with the univariate latent factor model for $\widetilde{Z}_{ijt}$.
Let $Y_{it}$ denote the latent factor at time $t$ for individual $i$.
The assumption of this model is that $Y_{it}$ is driving the multivariate response for each individual at each time point. 
That is, for each $i$, $j$, and $t$, we define  
\begin{equation}
\mu_{ijt} =  \beta_{0ij} + \beta_{1ij} Y_{it}
\label{eq:LFM}
\end{equation}
where $\beta_{0ij}$ is a metric-specific intercept term and $\beta_{1ij}$ is a metric-specific coefficient of the latent factor individual $i$.

We can extend (\ref{eq:LFM}) to an $M$-variate latent factor model where we now assume that the multivariate response might be a function of multiple latent factors. 
Let $Y_{1it}, \dots, Y_{Mit}$ denote the latent factors at time $t$ for individual $i$.
Then, we define $\mu_{ijt}$ as
\begin{equation}
\mu_{ijt} = \beta_{0ij} + \sum_{m=1}^M \beta_{mij} Y_{mit}
\end{equation}
where $\beta_{mij}$ captures the metric-specific effect of each latent factor.

We investigate the univariate and multivariate latent factors models in modeling the multivariate ordinal wellness data.
The two important covariates of interest identified above that are assumed to be driving athlete wellness include \emph{workload} and \emph{recovery}.
Therefore, we model the latent factors as functions of these variables. 
Specifically, we model the latent factors using distributed lag models such that we are able to capture the cumulative effects of workload and recovery on athlete wellness. 

Let $X_{1it}$ and $X_{2it}$ denote the workload and recovery variables for individual $i$ and time $t$, respectively. 
Starting with the univariate latent factor model, we model $Y_{it}$ as a linear combination of these lagged covariates. 
We write the distributed lag model for $Y_{it}$ as
\begin{equation}
Y_{it} = \sum_{l=0}^L \left(X_{1i,t-l}\alpha_{1il} + X_{2i,t-l}\alpha_{2il}\right) + \eta_{it}
\label{eq:Udl}
\end{equation}
where $\alpha_{1il}$ and $\alpha_{2il}$ are coefficients for the lagged $l$ covariates $X_{1i,t-l}$ and $X_{2i,t-l}$, and $\eta_{it}$ is an error term.
Here, we assume $\eta_{it} \sim N(0, \tau_i^2)$.
The distributed lag model is able to capture the covariate-specific cumulative effects at lags ranging from $l=0, \dots, L$. 
The benefit of the univariate approach is that latent factor $Y_{it}$ offers a univariate summary for individual $i$ on day $t$ as a function of both \emph{wellness} and \emph{recovery}. 
We can easily compare these univariate latent factors across days in order to identify anomalies in wellness across time for an individual.

The distributed lag model specification can also be utilized in the multivariate latent factor model. 
Having two important covariates of interest, we specify a bivariate latent factor model with factors $Y_{1it}$ and $Y_{2it}$.
Here, $Y_{1it}$ is modeled using a distributed lag model with covariate $X_{1it}$, and $Y_{2it}$ is modeled using a distributed lag model with covariate $X_{2it}$. 
The benefit of this approach is that we can infer about the separate metric-specific relationships with each of the lagged covariates for each individual.
That is, we can compare the relationships across metrics within an individual as well as within metric across individuals.
For $m =1, 2$, let
\begin{equation}
Y_{mit} = \sum_{l=0}^L X_{mi,t-l}'\alpha_{mil} + \eta_{mit}
\label{eq:Mdl}
\end{equation}
where $X_{mi,t}$, and $\alpha_{mil}$ are analogous to above, and $\eta_{mit}$ is the error term for factor $m$.
Again, we assume $\eta_{mit} \sim N(0,\tau^2_{mi})$.

It is worth mentioning that under certain parameter constraints, distributed lag models are equivalent to the ecological memory models proposed by \cite{Ogle2015}. 
That is, ecological memory models are a special case of distributed lag models where the lagged coefficients are assigned non-negative weights that sum to 1. 
For example, with $\text{E}(Y_{mit}) = \sum_{l=0}^L X_{mi,t-l}\alpha_{mil}$, the ecological memory model is such that $\alpha_{mil}>0$ and $\sum_{l=0}^L\alpha_{mil}=1$ for all $m$ and $i$.
Under this approach, $\boldsymbol{\alpha}_{mi} = (\alpha_{mi0}, \dots, \alpha_{miL})$ is modeled using a Dirichlet distribution.
The drawback of this approach is that this forces the relationship between the latent wellness metric $\widetilde{Z}_{ijt}$ and each element of the vector $(X_{mi0}, \dots, X_{miL})$ to be the same (e.g., all positive or all negative according to the sign of $\beta_{mij}$).
In our application, we desire the flexibility of having both positive and negative short- and long-term effects of training and recovery on athlete wellness.
For example, we might expect high-intensity training sessions to have immediate negative effects on wellness, but they could have positive impacts on wellness at longer time scales given proper recovery.


Under either the univariate or multivariate latent factor model, we can borrow strength across individuals by incorporating shared effects. 
Here, we include shared distributed lag coefficients. 
Recall that in (\ref{eq:Udl}) and (\ref{eq:Mdl}), $\alpha_{mil}$ denotes the lagged coefficient for variable $m$, individual $i$, and lag $l$.
We model $\alpha_{mil} \sim N(\alpha_{ml}, \psi_{ml})$ where $\alpha_{ml}$ is the global mean coefficient of covariate $m$ at lag $l$ and $\psi_{ml}$ represents the variability across individuals for this effect. 
We can obtain inference with respect to these global parameters to provide insight into the general effects of the covariates at various lags as well as the measures of variability across individuals.


\subsection{Identifiability constraints}
\label{sec:Ident}

We begin with a general depiction of the important identifiability constraints of the model parameters assuming one athlete (i.e., $n=1$). As such, we drop the dependence on $i$ in the following.
Additionally, we note that there is more than just one set of parameter constraints that will result in an identifiable model, and will therefore justify our choices with regard to desired inference when necessary. 

First, as is customary in probit regression models, the first threshold parameter, $\theta^{(1)}_{j} =0$ for each $j$ \citep{Chib1998}. 
This enables the identification of the intercept terms, $\beta_{0j}$.
Then, to identify the lag coefficients, $\alpha_{ml}$, for $m = 1, 2$, and $l = 0, \dots, L$, without loss of generality, we set the $\beta_{mj}$ factor coefficients for the first metric equal to 1 \citep{Cagnone2009}.
In the univariate latent factor model, this results in $\beta_{11} = 1$ and in the bivariate latent factor model, this results in $\beta_{11} = \beta_{21}=1$.
With the number of metrics $J>1$, we also must specify a common $\theta^{(2)}_1 = \cdots = \theta^{(2)}_J = \theta^{(2)}$ in order to identify the metric-specific latent factor coefficients, $\beta_{mj}$.

Another common identifiability constraint for probit regression models imposes a fixed variance for the latent continuous metrics, $\widetilde{Z}_{jt}$ \citep{Chib1998}. 
This is the variance of $\epsilon_{jt}$ from (\ref{eq:Ztilde}) which is denoted $\sigma_j^2$.
With metric specific threshold parameters $\theta^{(k)}_{j}$, we drop the dependence on $j$ such that $\sigma_1^2 = \cdots = \sigma_J^2 = \sigma^2$. 
One option is to fix $\sigma^2 = 1$ and model the variance parameters of the latent factors, $\tau^2$, in the univariate latent factor model, and $\tau^2_1$ and $\tau^2_2$ in the bivariate latent factor model \citep{Cagnone2009, Cagnone2018}.
However, since we are modeling the latent factors using distributed lag models, this approach can mask some of the effects of the lagged covariates as well as the relationships between the latent factors and the ordinal wellness metrics.
Therefore, we opt to work with the marginal variance of $\widetilde{Z}_{jt}$, which is equal to $$\text{Var}(\widetilde{Z}_{jt}) = \sigma^2 + \beta_{1j}^2 \tau^2$$ in the univariate factor model and 
$$\text{Var}(\widetilde{Z}_{jt}) = \sigma^2 + \beta_{1j}^2 \tau_1^2 + \beta_{2j}^2 \tau_2^2$$
in the bivariate factor model.
For $j=1$, this reduces to $\sigma^2 + \tau^2$ and $\sigma^2 + \tau_1^2 + \tau_2^2$.
We set $\sigma^2 + \tau^2 = 1$ and $\sigma^2 + \tau_1^2 + \tau_2^2 =1$ and use a Dirichlet prior with two and three categories, respectively. 
Details regarding this prior are given below.

In extending to modeling multiple athletes, we add subscript $i$ to each of the parameters and latent factors. 
That is, we have individual specific threshold parameters, intercepts, factor coefficients, latent factors, and variances. 
In addition, we introduce the global mean coefficients, $\alpha_{ml}$, and variances, $\psi_{ml}$.
By imposing the same set of constraints above for each individual, the model parameters are identifiable.

Fitting this model to the referee ordinal wellness data discussed in Section \ref{Sec:Data} requires one additional modification.
In looking at the ordinal response distributions (Figures 1 and 11), notice that for some individuals and some metrics, some ordinal values have few, if any, counts (e.g., Athlete A: Stress). 
In such a case, there is no information in the data to inform about the cut points for these individual and metric combinations. 
Therefore, we drop the individual specific threshold parameters to leverage information across athletes for each metric.
With this modification, we can relax the constraint on $\theta^{(2)}$ to allow for metric specific thresholds, $\theta^{(2)}_1, \dots, \theta^{(2)}_J$. 
The shared metric-specific threshold approach across individuals is preferred over having individual threshold parameters that are shared across metrics for two reasons.
First, some referees have very small counts for some ordinal values, even when aggregated across metrics, resulting in challenges in estimating these parameters. When aggregating across referees for a given metric, the distribution of observations across ordinal values is much more uniform.
Second, by retaining the metric-specific threshold parameters, we can more easily compare the metric-specific relationships with the latent factor(s). 
That is, we can directly compute correlations between the latent factor(s) and the latent continuous wellness metrics as discussed below. 
Due to ordinal data not having an identifiable scale, specifying individual threshold parameters that are shared across metrics requires computing more complex functions of the model parameters in order to obtain this important inference.

\subsection{Model inference and priors}
\label{sec:Inf1}

Model inference was obtained in a Bayesian framework. 
Prior distributions are assigned to each model parameter and non-informative and conjugate priors were chosen when available. 
Each global mean lagged coefficient parameter is assigned an independent, conjugate hyper prior where $\alpha_{ml} \sim N(0,10)$ for all $m=1, 2$ and $l = 1, \dots, L$. 
The variance parameters are assigned independent Inverse-Gamma(0.01, 0.01) priors.
The latent factor coefficients are assigned independent normal priors, where $\beta_{mij} \sim N(0,10)$ for $m=0, 1$ in the univariate factor model and $m=0, 1, 2$ in the bivariate factor model.

Given the identifiability constraints above for the variance parameters, we specify a two category Dirichlet prior for $(\sigma^2, \tau^2)$ in the univariate latent factor model and a three category Dirichlet prior for $(\sigma^2, \tau_1^2, \tau_2^2)$ in the bivariate model. 
Both Dirichlet priors are defined with concentration parameter 10 for each category.

The threshold parameters were modeled on a transformed scale due to their order restriction where $\theta_{ij}^{(k-1)} \leq \theta_{ij}^{(k)}$.
To ensure these inequalities hold true, with $\theta_{ij}^{(1)} = 0$ for all $i$ and $j$, we define $\widetilde{\theta}_{ij}^{(k)} = \text{log} (\theta_{ij}^{(k)} - \theta_{ij}^{(k-1)})$ for $k=2, 3, 4$ and model $\widetilde{\theta}_{ij}^{(k)} \stackrel{iid}{\sim} N(0, 1)$.
This transformation improves mixing and convergence when using MCMC for model inference \citep{Higgs2010}.
Sampling the threshold parameters requires a Metropolis step within the MCMC algorithm.

\subsection{Posterior inference}
\label{sec:Inf2}
Important posterior inference includes estimates of the model parameters as well as correlation and relative importance measures for each metric.
Dropping the dependence on $i$ for ease of notation, let $C_j$ define the correlation between latent ordinal response metric $\widetilde{\mathbf{Z}}_j = (\widetilde{Z}_{j1}, \dots, \widetilde{Z}_{jT})'$ and the univariate latent factor $\mathbf{Y} = (Y_1, \dots, Y_T)'$, computed as 
\begin{equation}
C_j = \text{corr}(\widetilde{\mathbf{Z}}_j, \mathbf{Y}).
\label{eq:CorrU}
\end{equation}
For the multivariate latent factor model, we can define analogous correlations for each factor $m$ where
\begin{equation}
C_{jm} = \text{corr}(\widetilde{\mathbf{Z}}_j, \mathbf{Y}_m).
\label{eq:CorrM}
\end{equation}
These correlations provide a measure for which to compare the importance of each wellness metric in capturing the variation in the latent factor. 
To compare across metric, we compute the relative importance of each metric for the univariate latent factor model as
\begin{equation}
R_j = \frac{|C_j|}{\sum_{j'=1}^J |C_{j'}|}
\label{eq:RIU}
\end{equation}
and as 
\begin{equation}
R_{jm} = \frac{|C_{jm}|}{\sum_{j'=1}^J |C_{j'm}|}
\label{eq:RIM}
\end{equation}
for the multivariate factor model.
Metrics with higher relative importance indicate that they are more important in capturing the variation in the latent factor.
For the univariate latent factor model, these relative importance scores could be used as weights in computing an overall wellness score for each individual. 
Then, these latent factors could be monitored through time to identify possible changes in each athlete's wellness in response to training and recovery throughout the season.
For example, we can investigate the variation in the latent factors for each athlete by comparing match days to the days leading up to and following the match.
Similarly, for the multivariate metric, these weights could identify which metrics are more or less important in explaining the variation in each particular latent factor, and again, can be monitored throughout the season to identify potential spikes in wellness as a response to training and recovery.
We can obtain full posterior distributions, including estimates of uncertainty, of all correlations and relative importance metrics post model fitting.

\section{Application: modeling athlete wellness}
\label{Sec:Results}
We apply our model to the subjective ordinal wellness data and workload and recovery data for 20 MLS referees collected during the 2015 and 2016 seasons.
We investigated the effects of the workload and recovery variables on wellness for lags up to 10 day. 
Therefore, we limited the analysis to daily data for which at least 10 prior days of workload and recovery data were available. 
The number of observations for each individual ranged from 170 to 467 days.

The univariate and multivariate latent factor models were each fitted to the data.
Model inference was obtained using Markov chain Monte Carlo and a hybrid Metropolis-within-Gibbs sampling algorithm. 
The chain was run for 100,000 iterations, and the first 20,000 were discarded as burn-in.
Traceplots of the chain for each parameter were investigated for convergences and no issues were detected.

\begin{figure}
\begin{center}
\includegraphics[scale=.6]{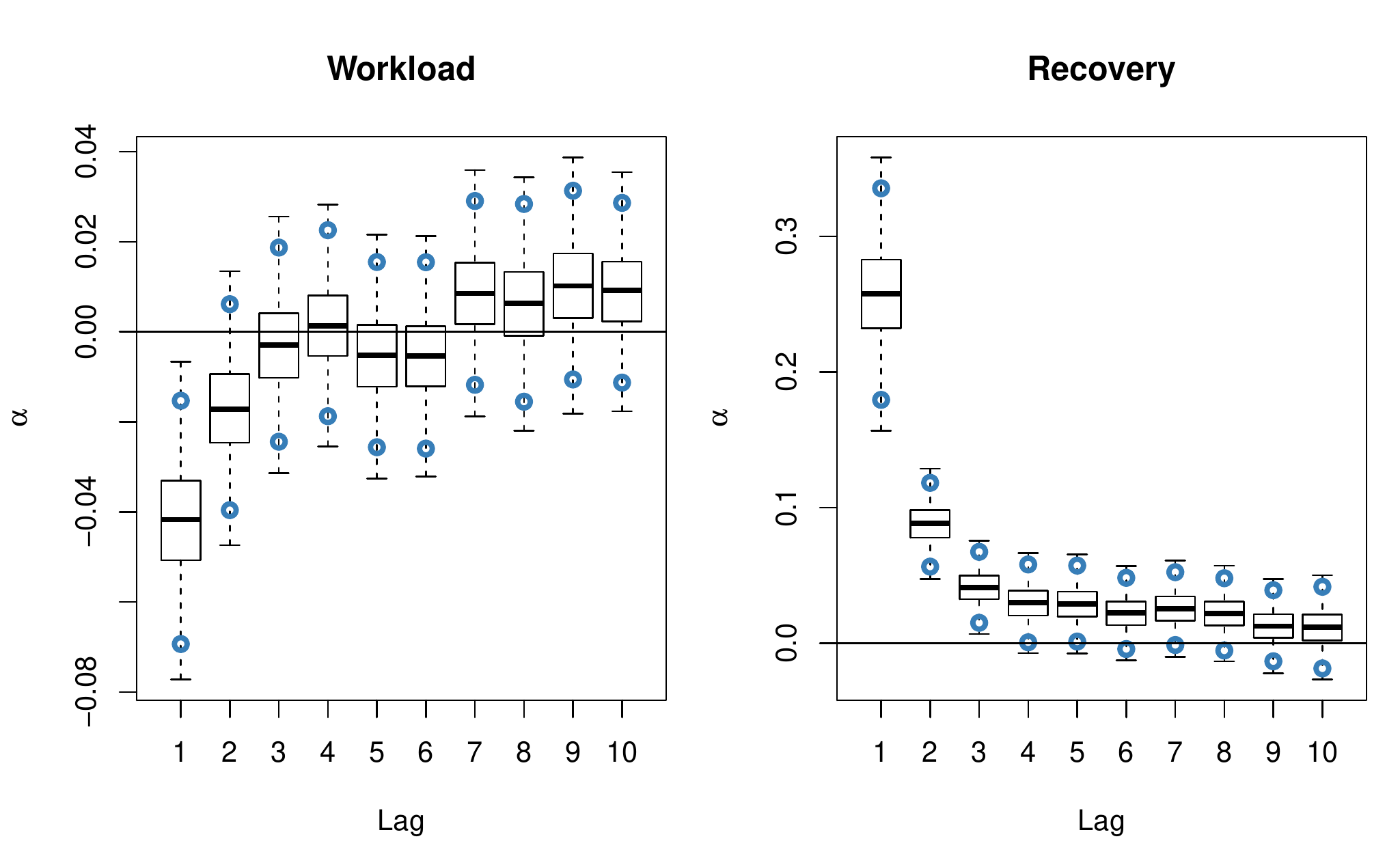}
\caption{Distribution of the global lagged coefficients for the one factor model for workload (left) and recovery (right). {\color{myblue}$\boldsymbol{\circ}$} indicates 95\% credible interval. \label{Fig:AlphaGlobU}}
\end{center}
\end{figure}

Boxplots of the posterior distributions of the global lagged coefficients of the univariate latent factor model are shown in Figure \ref{Fig:AlphaGlobU} for the workload and recovery covariates.
Also shown are the upper and lower limits of the central 95\% credible intervals.
In general, workload is negatively related with athlete wellness, and the previous day's workout (lag equal to 1) is the most significant. 
This implies that, in general, workload has an acute effect on player wellness, such that a heavy workload on the previous day tends to lead to a decrease in wellness on the following day.
The lagged coefficients of the recovery variable show a positive relationship between recovery and wellness, where a large recovery value corresponds to high sleep quality and quantity.
The lagged coefficients for this variable are significant for lags 1 through 5 as indicated by the 95\% credible intervals not including 0.
These significant lagged coefficients suggest that sleep quality and quantity may have a longer lasting effect on athlete wellness. 

\begin{figure}
\begin{center}
\includegraphics[scale=.37]{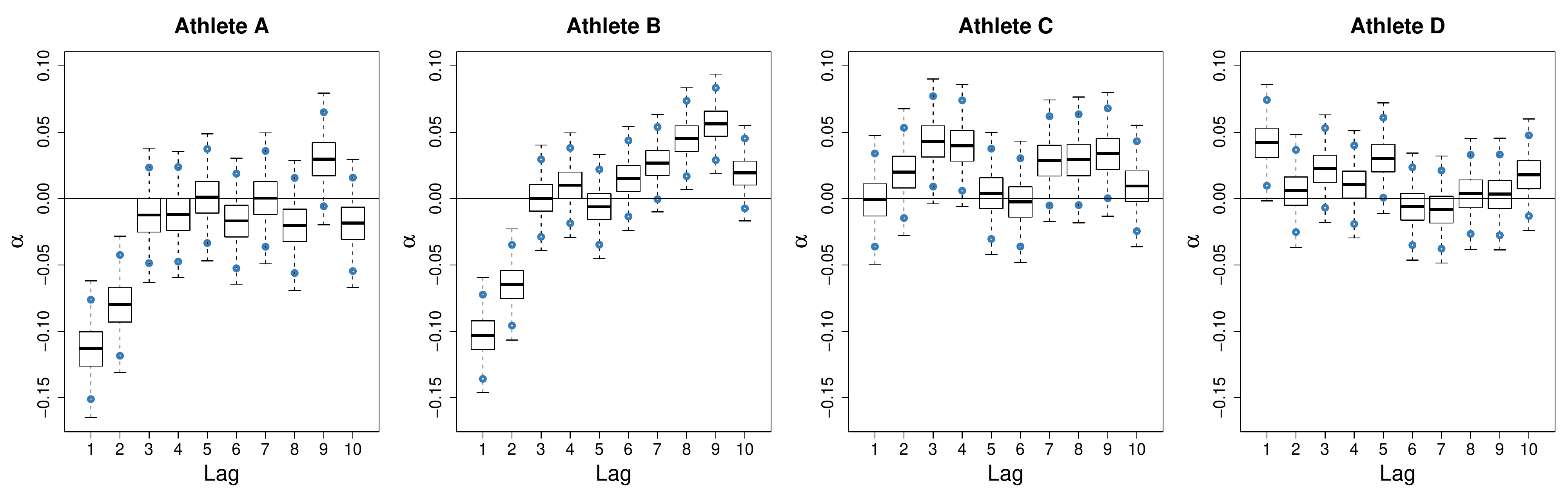}
\caption{Distribution of the individual specific lagged coefficients for the workload covariate in the univariate latent factor model. {\color{myblue}$\boldsymbol{\circ}$} indicates 95\% credible interval.\label{Fig:AlphaIndUW}}
\end{center}
\end{figure}
\begin{figure}
\begin{center}
\includegraphics[scale=.37]{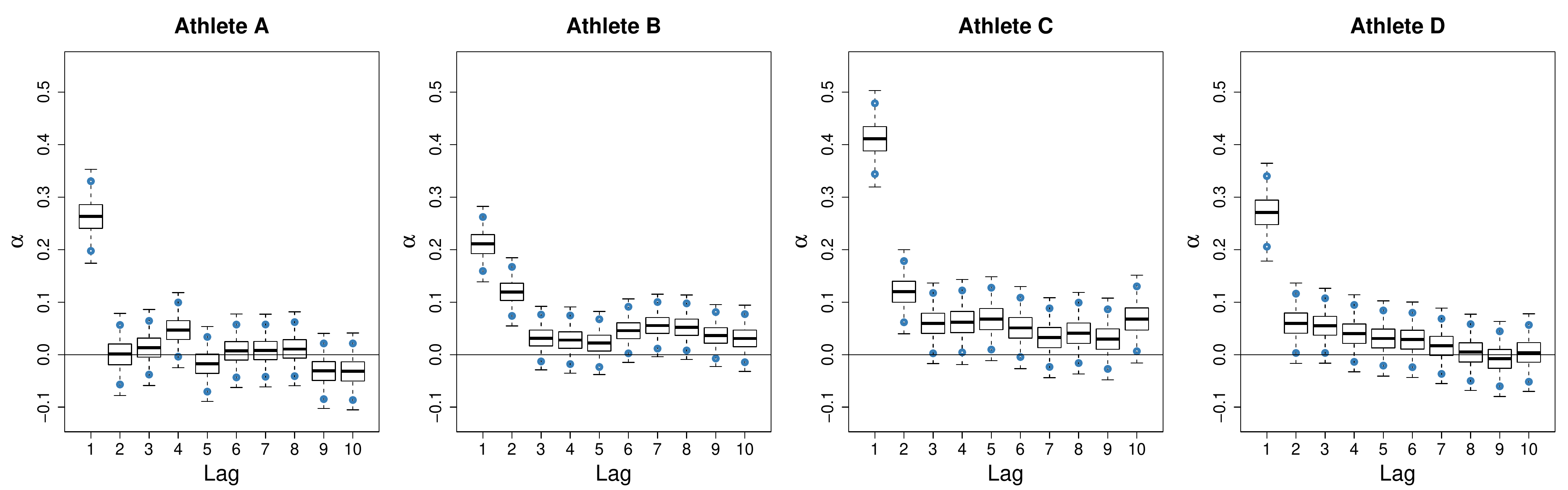}
\caption{Distribution of the individual specific lagged coefficients for the recovery covariate in the univariate latent factor model. {\color{myblue}$\boldsymbol{\circ}$} indicates 95\% credible interval.\label{Fig:AlphaIndUR}}
\end{center}
\end{figure}

Posterior distributions of the individual-specific lagged coefficients are shown for Athlete A, B, C, and D in Figures \ref{Fig:AlphaIndUW} and \ref{Fig:AlphaIndUR} for the workload and recovery variables, respectively. 
In general, there is a lot of variation between individuals with regard to the lagged effects of the two variables.
For example, Athlete A and B experience significant negative effects of workload at both lags 1 and 2, whereas Athlete C and D do not experience such negative effects. 
In fact, a heavy workload on the previous day has a positive relationship with wellness for Athlete D.
For all four athletes, we see positive effects of workload at longer lags (e.g., lag 9 for Athlete A, lags 7-9 for Athlete B). 
The individual-specific lagged coefficients of the recovery variable show that the previous nights sleep quantity and quality have a very significant positive relationship with wellness for each athlete (Figure \ref{Fig:AlphaIndUR}).
However, we detect a more short-term effect of recovery for Athlete A and B (2 days) relative to Athlete C and D (3+ days) than the average shown in Figure \ref{Fig:AlphaGlobU}.

%

\begin{figure}
\begin{center}
\includegraphics[scale=.48]{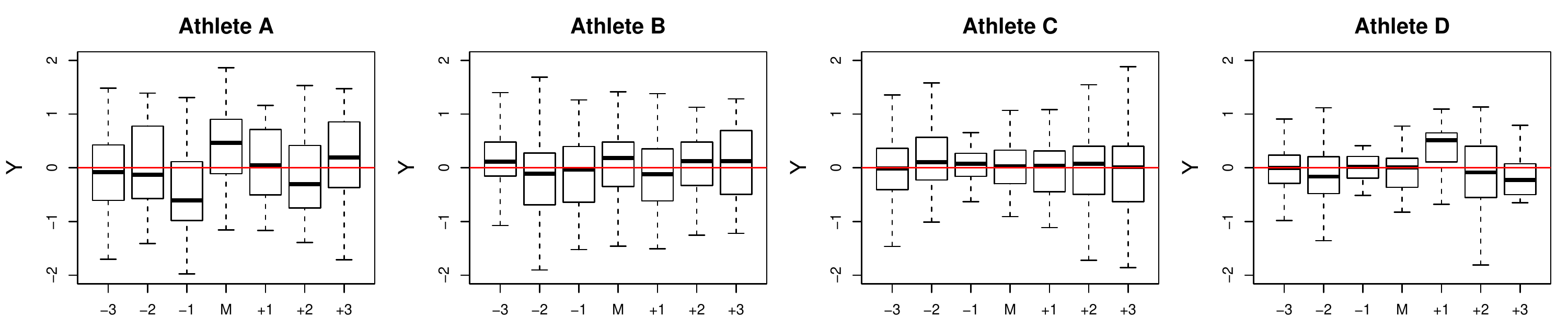}
\caption{Posterior mean estimates of the latent factor, $Y_{it}$, for each individual and match (M), as well as the three days leading up to (-3, -2, -1) and following (+1, +2, +3) the match. The estimates shown are mean-centered for each match. Athlete A, B, C, and D reffed 30, 41, 38, and 19 matches, respectively during the 2015 and 2016 seasons. \label{Fig:LatY}}
\end{center}
\end{figure}

The latent factors, $Y_{it}$, provide a univariate measure of wellness for each individual on each day. 
Figure \ref{Fig:LatY} shows boxplots of the posterior mean estimates of $Y_{it}$ for the four athletes for match days, denoted ``M," compared to the 3 days leading up to and following the match. 
Due to possible variation throughout the seasons, the estimates are centered within each match week by subtracting the 7-day average. 
Estimates of the latent factors vary throughout the 7-day period for each athlete. 
In general, the wellness of Athlete A is highest on match day relative to the days leading up to and following the match.
The wellness estimates for Athlete B and C show less variation across days, although wellness for Athlete B is lower, on average, the day following the match relative to the match day.
Wellness for Athlete D appears similar across all days except for the day following the match, in which wellness is higher.

\begin{figure}
\begin{center}
\includegraphics[scale=.4]{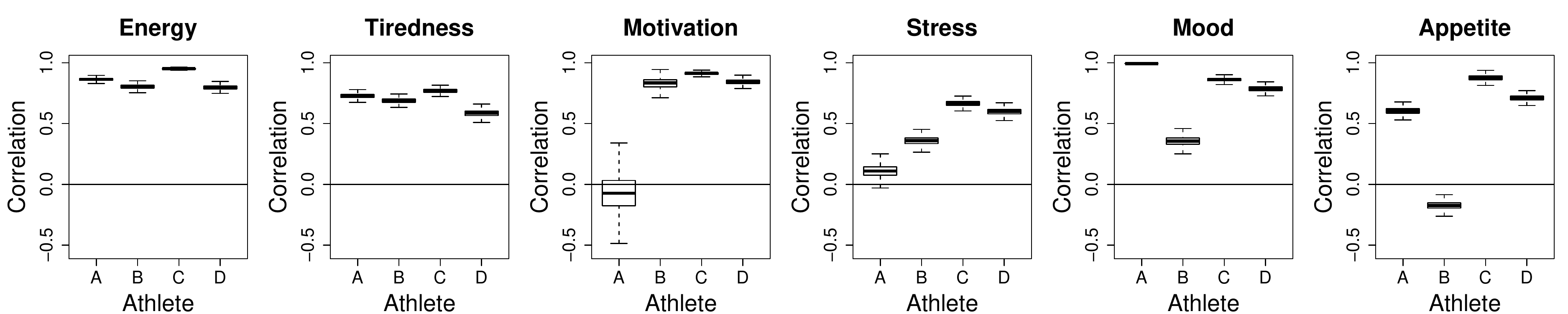}
\caption{Correlation between the univariate latent factor, $\mathbf{Y}_i$ and $\widetilde{\mathbf{Z}}_{ij}$ for each athlete and latent metric variable. \label{Fig:CorrU}}
\end{center}
\end{figure}

We computed the correlation between the vectors of the latent continuous response metric, $\widetilde{\mathbf{Z}}_{ij}$ and the univariate latent factor, $\mathbf{Y}_{i}$, for each athlete and metric. 
Boxplots of the posterior distributions for these correlations for the four athletes are shown in Figure \ref{Fig:CorrU}, indicating variation both within metric across individuals and across metrics within individual. 
The majority of the significant correlations between the ordinal wellness metric and latent factor are positive, although the correlation was negative for Athlete B for the appetite metric. 

\begin{figure}
\begin{center}
\includegraphics[trim={4cm 0 4cm 0},clip,width=.5\textwidth]{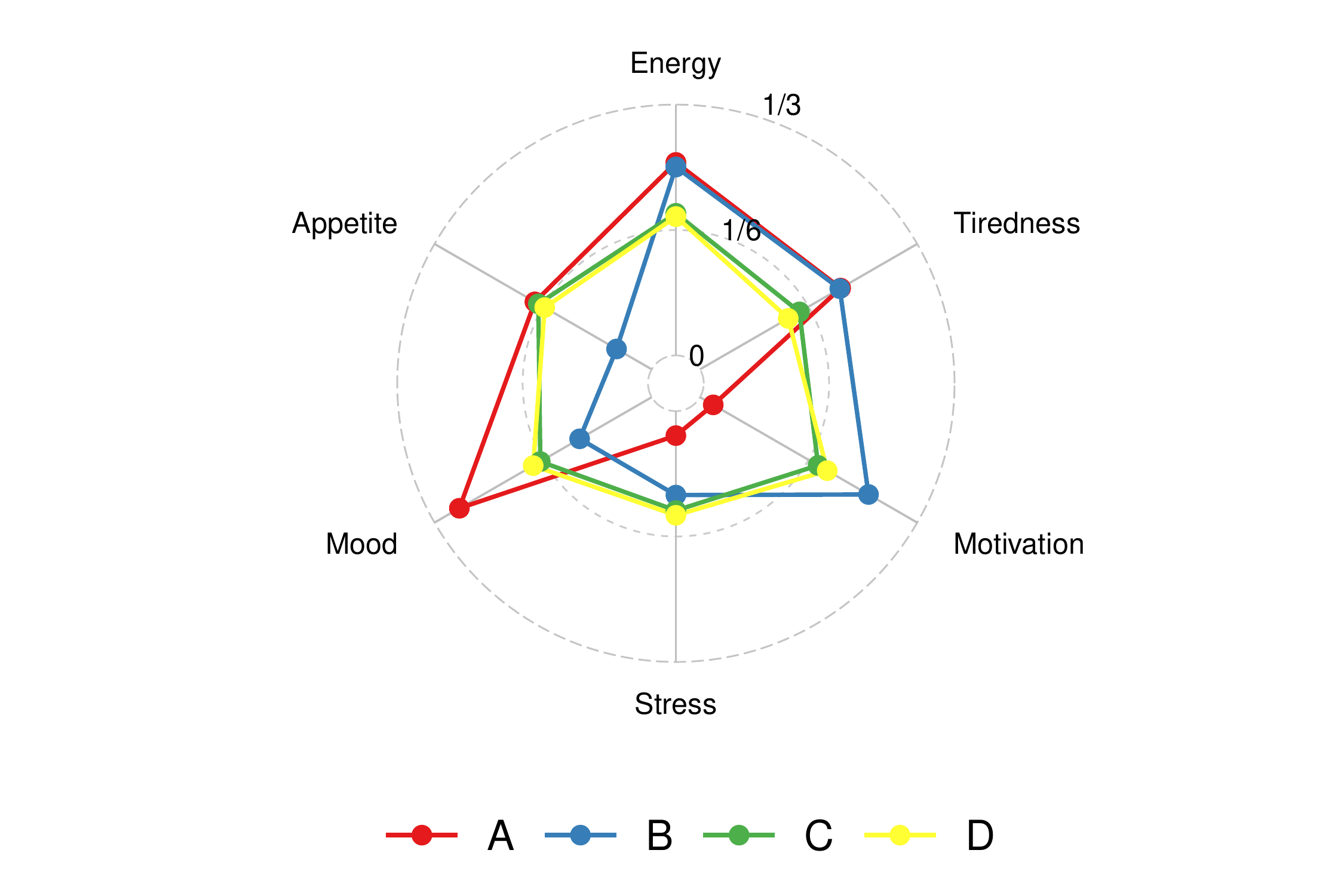}
\caption{Posterior mean estimates of the relative importance statistics, $R_j$, defined in (\ref{eq:RIU}) for each athlete and metric. \label{Fig:RIU} }
\end{center}
\end{figure}

To compare the significance of the different metrics within an individual in relation to the latent factor, we compute the relative importance statistics defined in (\ref{eq:RIU}).
The relative importance statistics give a measure of the ability of each wellness metric at capturing the variation in the latent factor. 
The posterior mean estimates of $R_j$ for each of the four athletes across the six metrics are shown in Figure \ref{Fig:RIU}.
The relative importance of each ordinal wellness metric varies across the athletes. 
Note that a value of 1/6 for each metric would correspond to an equal weighting.
The most notable similarity between the four athletes is the high relative importance of \emph{energy}, with each greater than 1/6.
The relative importance of \emph{motivation} and \emph{mood} vary a lot between athletes.
The estimates for Athlete C and D closely resemble an equal weighting scheme across the six metrics, whereas A and B each have unequal relative importance estimates with emphasis on mood, energy and tiredness for Athlete A, and motivation, energy, and tiredness for Athlete B. These results clearly depict a difference between computing the average across all metrics and the utility of the multivariate model in leveraging the individual wellness measures.
Plots of the correlation and relative importance metrics for all 20 athletes are included in Figures \ref{Fig:CorrsUALL} and \ref{Fig:RIUALL} of the Supplementary Material.

%

\begin{figure}
\begin{center}
\includegraphics[scale=.4]{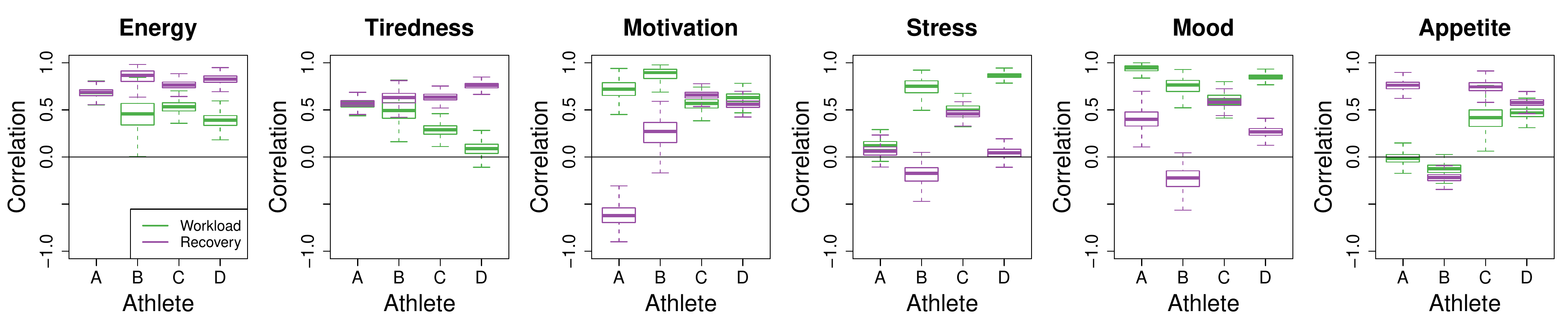}
\caption{Correlation between each latent continuous wellness metric, $\widetilde{\mathbf{Z}}_{ij}$ and workload latent factor, $\mathbf{Y}_{i1}$ (left), and recovery latent factor, $\mathbf{Y}_{i2}$ (right) for each athlete $i$ and latent metric variable $j$. \label{Fig:CorrM}}
\end{center}
\end{figure}

The multivariate latent factor model resulted in similar global and individual lagged coefficient estimates as the univariate model. (See Figures \ref{Fig:M1} - \ref{Fig:M3} of the Supplementary Material). 
In addition, the variation in the estimates of the two latent factors across days leading up to and following each match also appeared similar to the univariate model (Figures \ref{Fig:LatY1} and \ref{Fig:LatY2}). 
Important inference from the multivariate model consists of the factor-specific correlations and relative importance estimates for each wellness metric. 
Posterior distributions of the correlation estimates are shown in Figure \ref{Fig:CorrM} for both workload and recovery variables for the same four athletes, Athlete A, B, C, and D. (A similar figure with all athletes is given in Figure \ref{Fig:CorrMALL} of the Supplementary Material). 
This figure shows some important similarities and differences for each of the wellness metrics in terms of the correlations with the two latent factors. 
Both energy and tiredness show stronger correlations with recovery than workload for Athlete B, C, and D. 
The motivation metric is significantly more correlated with workload than recovery for Athlete A and B, whereas it is similar for Athlete C and D.
Stress and mood both appear more strongly correlated with workload, whereas appetite appears more strongly correlated with recovery for each athlete.

\begin{figure}
\begin{center}
\includegraphics[trim={4cm 0 4cm 0},clip,width=.5\textwidth]{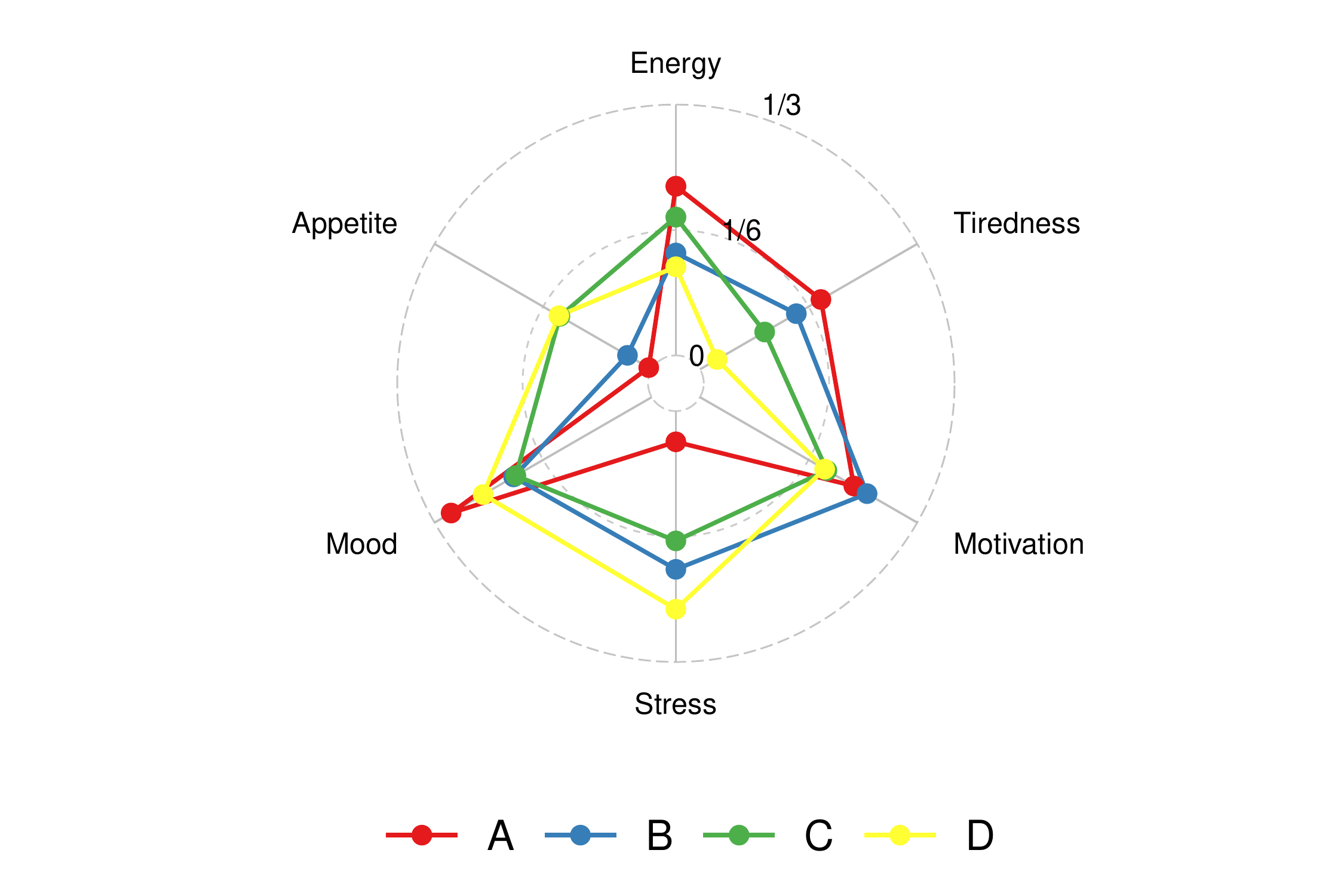}~~
\includegraphics[trim={4cm 0 4cm 0},clip,width=.5\textwidth]{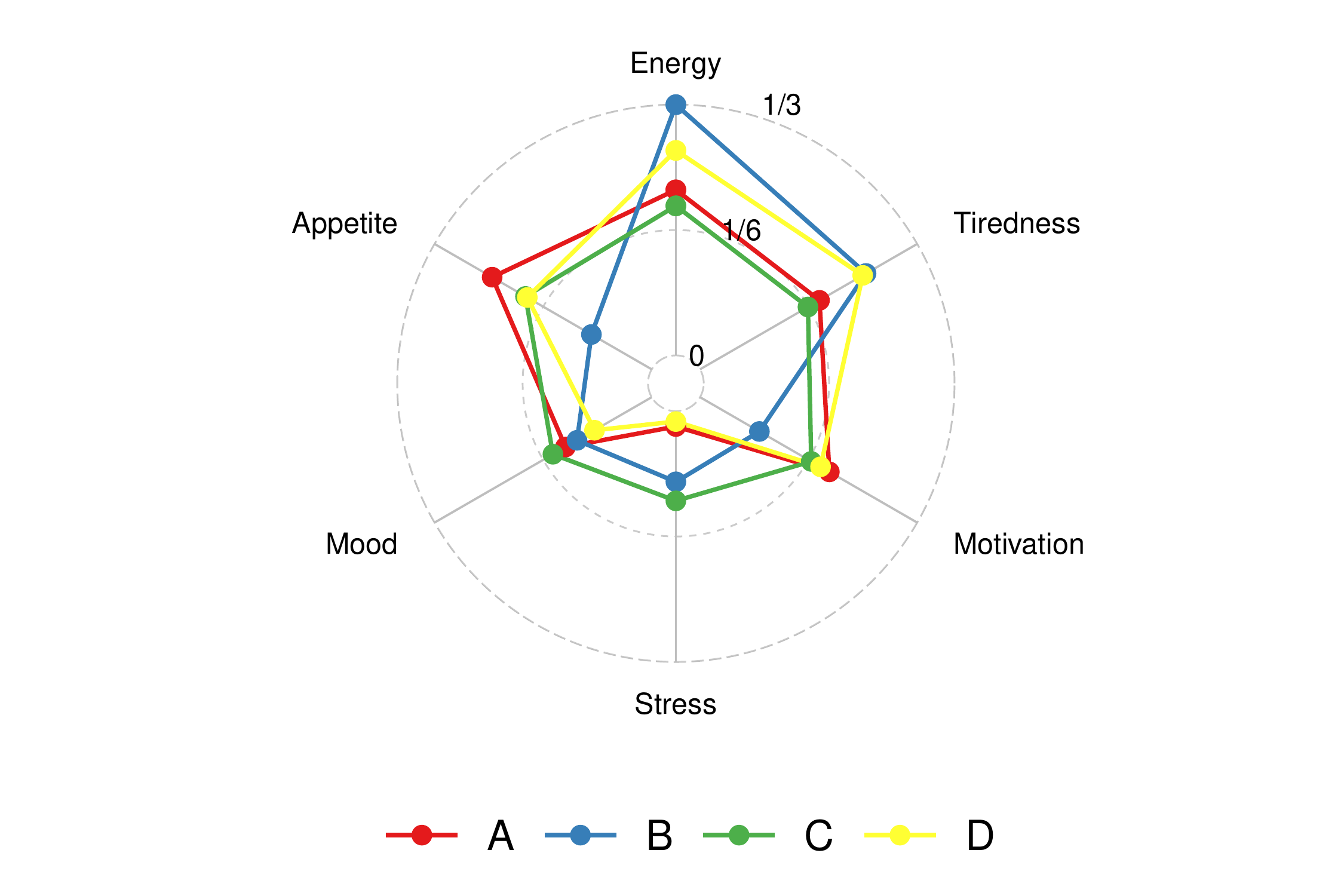}
\caption{Posterior mean estimates of the relative importance statistics, $R_{mj}$, defined in (\ref{eq:RIM}) for each athlete and metric for the workload latent factor (left) and recovery latent factor (right).\label{Fig:RIM}}
\end{center}
\end{figure}

Posterior mean estimates of relative importance for each latent factor for the four athletes are shown in Figure \ref{Fig:RIM}.
Some wellness metrics that appeared insignificant in the univariate latent factor model now show significance when the workload and recovery latent factors are considered separately. 
For example, motivation appears significant for both workload and recovery for Athlete A whereas it was the least important metric in the univariate latent model.
The relative importance of mood on the workload latent factor is greater than 1/6 for each athlete, and is the highest relative importance for Athlete A.
Energy and tiredness appear to capture the majority of the variation in the recovery latent factor for Athlete B.
Interestingly, Athlete C retains a fairly equal weighting scheme across the six metrics for both workload and recovery.
The relative importance of stress is high for workload and low for recovery for Athlete D, whereas tiredness is low for workload and high for recovery. Additional comparisons between all athletes with respect to the relative importance for each metric and latent factor can be made looking at Figure \ref{Fig:RIMALL} of the Supplementary Material. Interestingly, none of the metrics appear to be uniformly insignificant across the 20 athletes, providing justification in each component of the self-assessment survey.

%
%

\section{Discussion}
\label{Sec:Disc}
We develop a joint multivariate latent factor model to study the relationships between athlete wellness, training, and recovery using subjective and objective measures. 
Importantly, the multivariate response model incorporates the information from each of the subjective ordinal wellness variables for each individual.
Additionally, the univariate and bivariate latent factors are modeled using distributed lag models to identify the short- and long-term effects of training and recovery. 
Individual-specific parameters enable individual-level inference with respect to these effects. 
The relative importance indices provide individual-specific estimates of the sensitivity of each ordinal wellness metric to the variation in training and recovery.
The joint modeling approach enables the sharing of information across individuals to strengthen the results.

We applied our model to daily wellness, training, and recovery data collected across two MLS seasons. While the results show important similarities and differences across athletes with regard to the training and recovery effects on wellness and the importance of each of the wellness variables, the model could provide new and insightful information when applied to competing athletes. When referencing physical performance, our findings align with important known differences in training programs for referees and players. Training programs aim at maximizing the physical performance of players on match days, whereas referee training does not place the same significance on these days. This suggests interesting comparisons could be made using the results of this type of analysis between athletes who compete in different sports with differing levels of intensity and periods of recovery. For example, football has regular weekly game schedules at the college and professional levels, whereas soccer matches are scheduled typically twice per week, and hockey leagues often play two and three-game series with games on back-to-back nights. Maximizing player performance under these different competition schedules and levels of intensity using subjective wellness data is an open area for future work.

Distributed lag models, like those applied in this work, make the assumption that the lagged coefficients are constant in time. 
That is, the effect of variable $X_{1,t-l}$ at lag $l$ on the response at time $t$ is captured by $\alpha_{1l}$ and is the same for all $t$. 
In terms of training and recovery for athletes, one could argue that these effects might vary throughout a season as a function of fitness and fatigue. 
For example, if an athlete's fitness level is low during the early part of the season, a hard training session might have a longer lasting effect on wellness than it would mid-season when the athlete is at peak fitness. Alternatively, as the season wears on, an athlete might require a longer recovery time in order to return to their maximum athletic performance potential. As future work, we plan to incorporate time-varying parameters into the distributed lag models. This will require strategic model development in
in order to minimize the number of additional parameters and retain computation efficiency in model fitting.
The scope of this future work spans beyond sports, as the lagged effects of environmental processes could also have important time-varying features. 




\bibliographystyle{apalike}
\bibliography{LFMreferee}


\newpage
\FloatBarrier

\FloatBarrier

\appendix

\renewcommand{\thefigure}{A\arabic{figure}}

\setcounter{figure}{0}


\section{Supplementary Material}

\begin{figure}[h]
\begin{center}
\includegraphics[scale=.5]{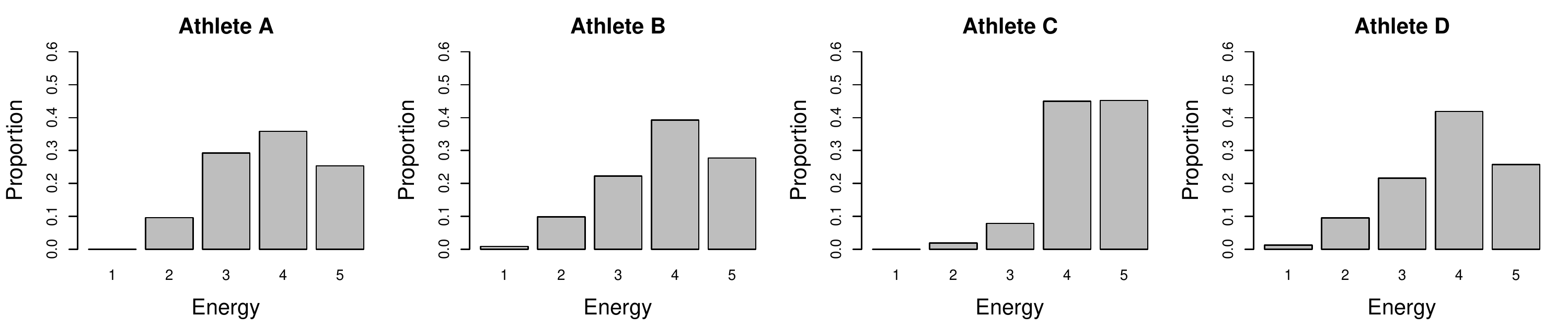}
\includegraphics[scale=.5]{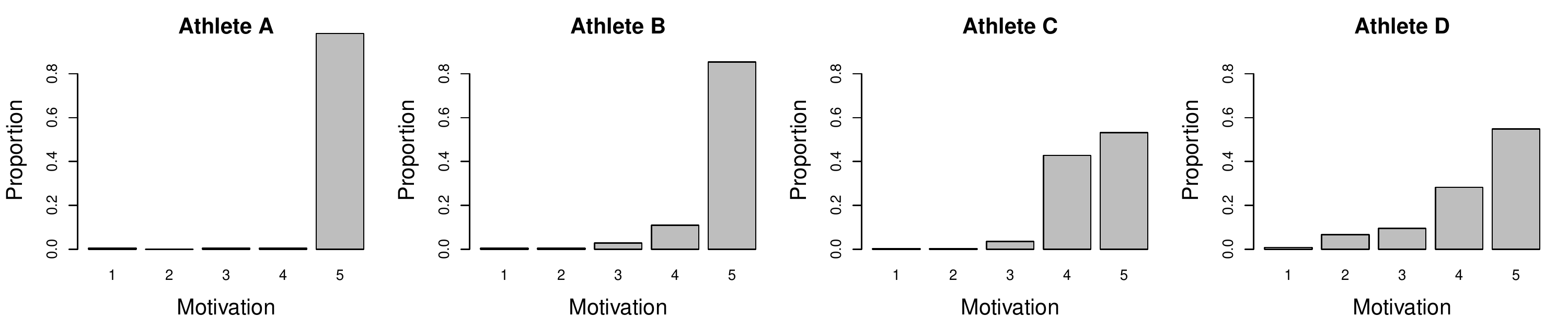}
\includegraphics[scale=.5]{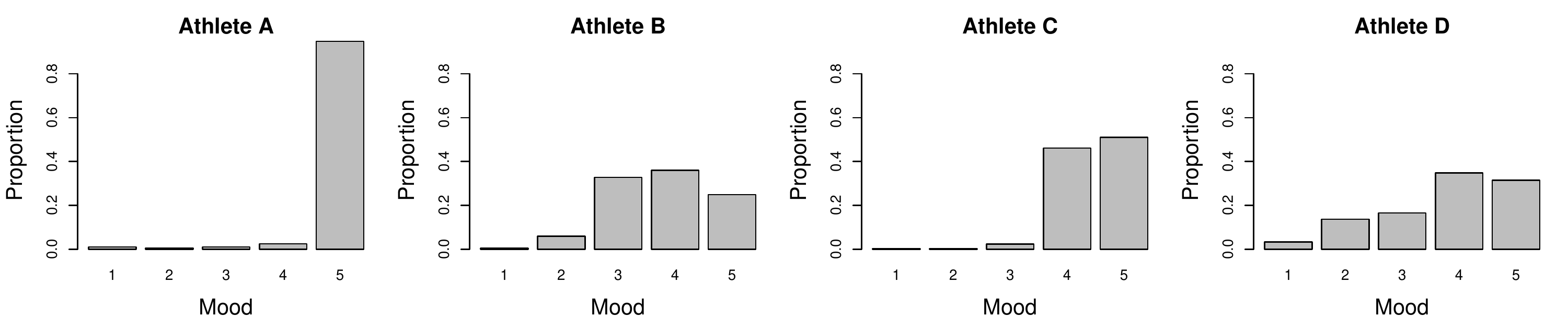}
\includegraphics[scale=.5]{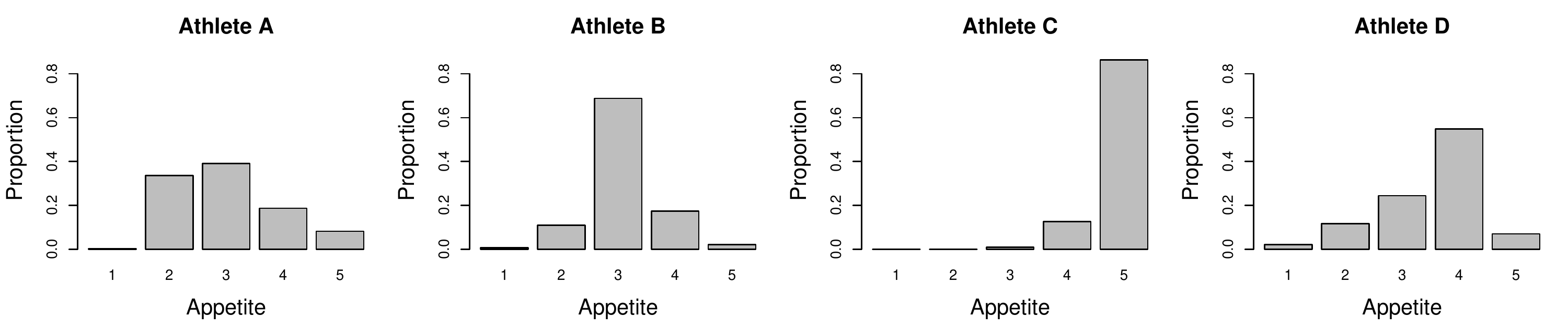}
\caption{Raw summaries of the ordinal wellness metrics \emph{energy}, \emph{motivation}, \emph{mood}, and \emph{appetite} for four athletes. \label{Fig: WM2}}
\end{center}
\end{figure}

\begin{figure}
\begin{center}
\includegraphics[scale=.4]{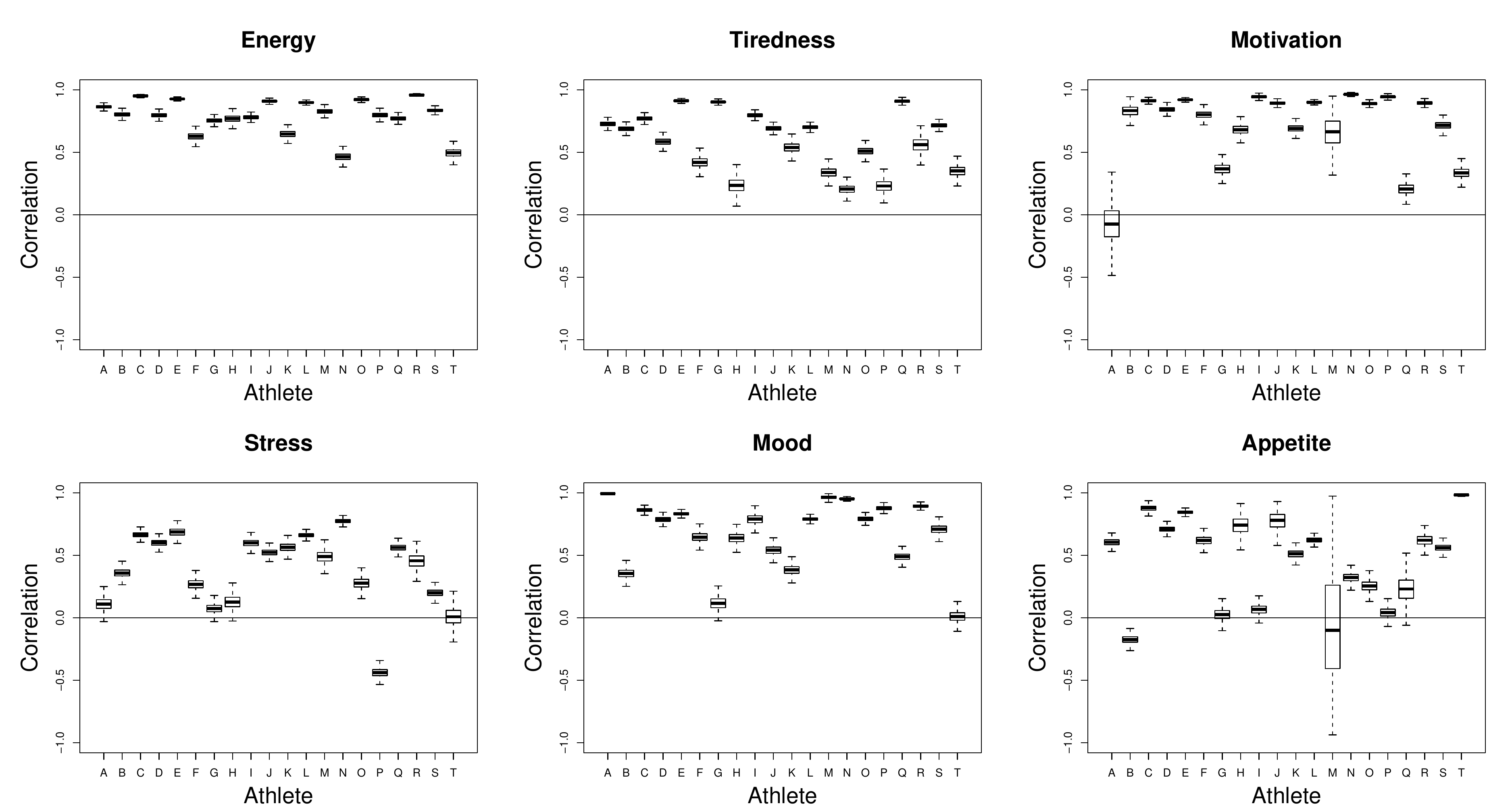}
\caption{Boxplots of the posterior distributions of the correlation between the univariate latent factor, $\mathbf{Y}_i$ and $\widetilde{\mathbf{Z}}_{ij}$, for all athletes and metric. \label{Fig:CorrsUALL}}
\end{center}
\end{figure}

\begin{figure}
\begin{center}
\includegraphics[scale=.4]{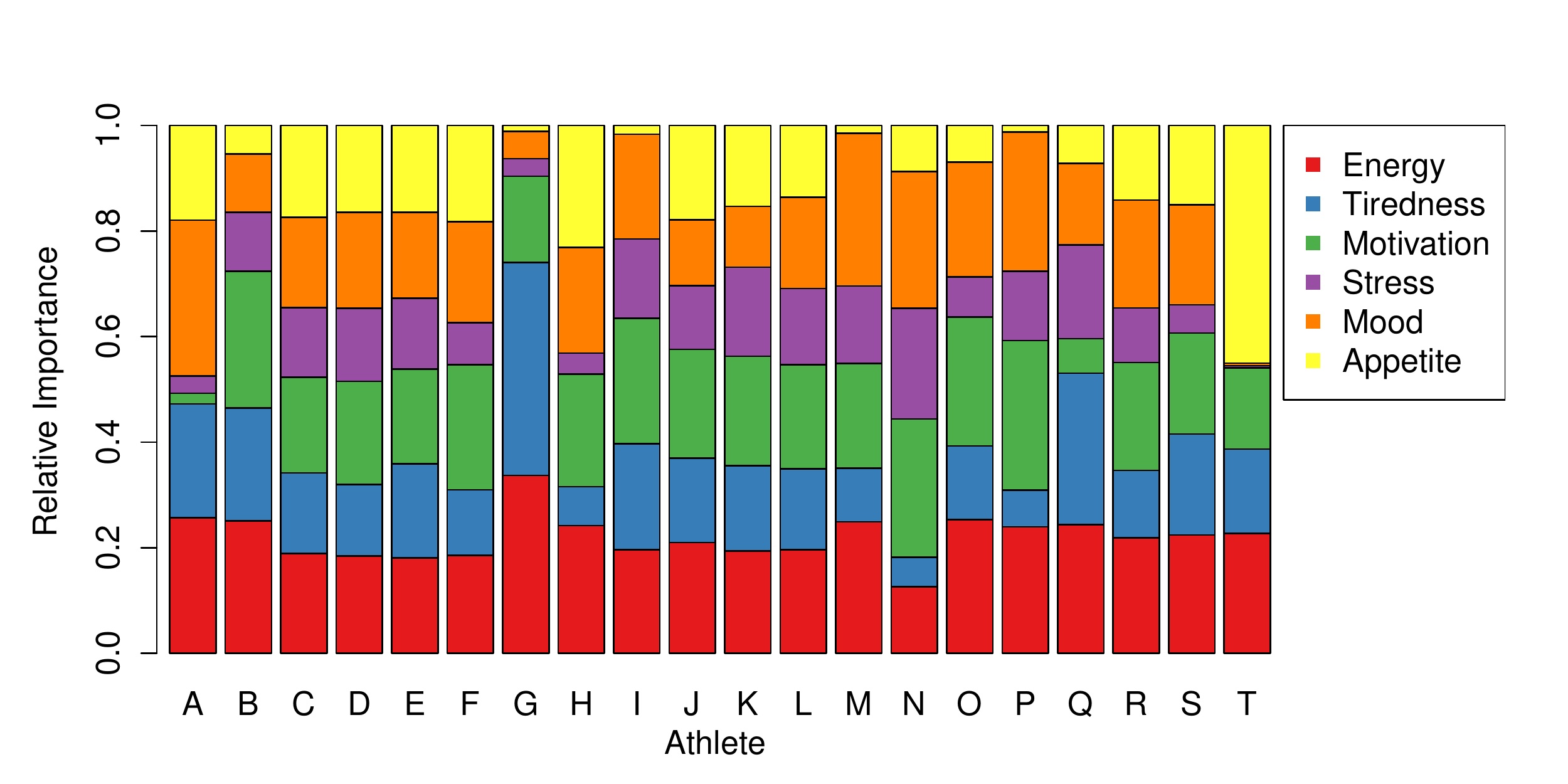}
\caption{Posterior mean estimates of the relative importance statistics, $R_j$, defined in (9) for all athletes and metric.\label{Fig:RIUALL}}
\end{center}
\end{figure}

\begin{figure}
\begin{center}
\includegraphics[scale=.6]{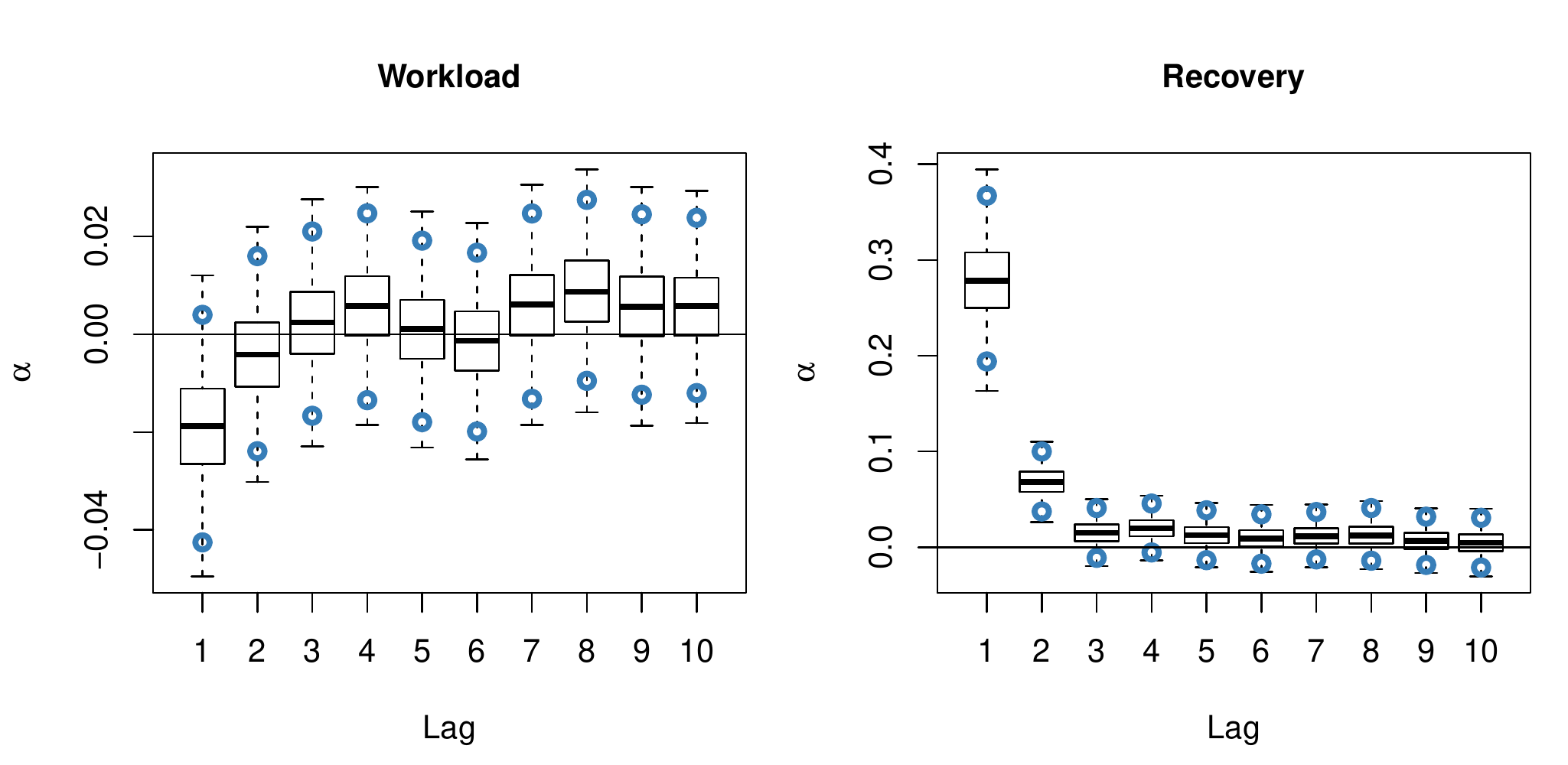}
\caption{Distribution of the global lagged coefficients for the two factor model for workload (left) and recovery (right). {\color{myblue}$\boldsymbol{\circ}$} indicates 95\% credible interval. \label{Fig:M1}}
\end{center}
\end{figure}

\begin{figure}
\begin{center}
\includegraphics[scale=.37]{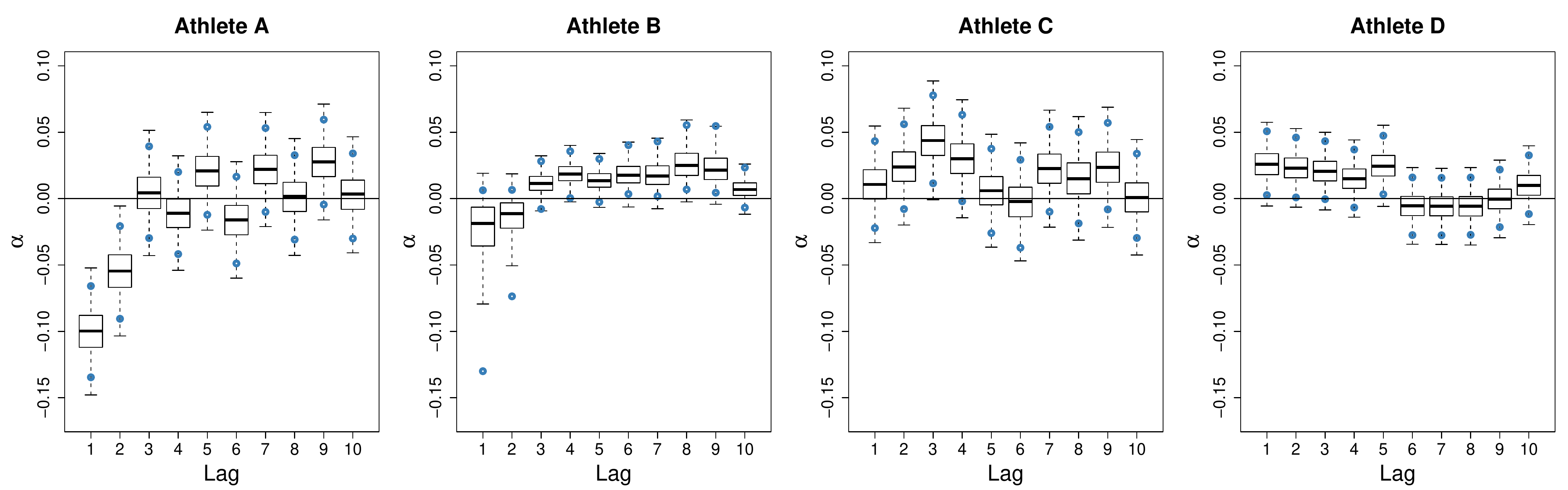}
\caption{Distribution of the individual specific lagged coefficients for the workload latent factor. {\color{myblue}$\boldsymbol{\circ}$} indicates 95\% credible interval. \label{Fig:M2}}
\end{center}
\end{figure}

\begin{figure}
\begin{center}
\includegraphics[scale=.37]{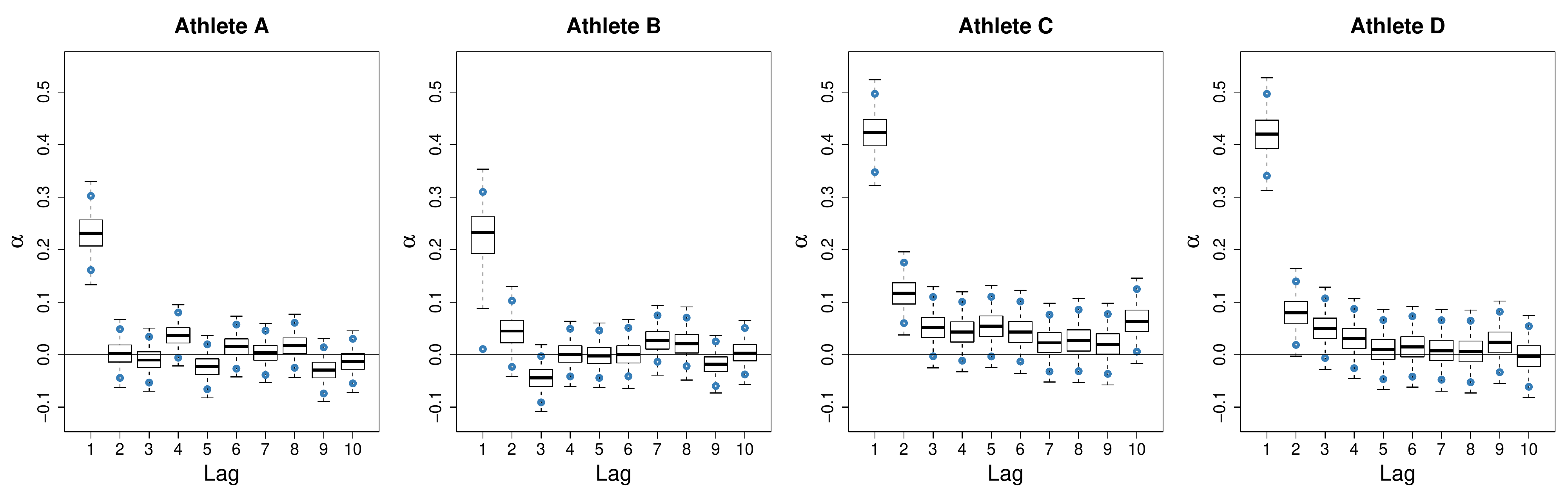}
\caption{Distribution of the individual specific lagged coefficients for the recovery latent factor. {\color{myblue}$\boldsymbol{\circ}$} indicates 95\% credible interval. \label{Fig:M3}}
\end{center}
\end{figure}

\begin{figure}
\begin{center}
\includegraphics[scale=.48]{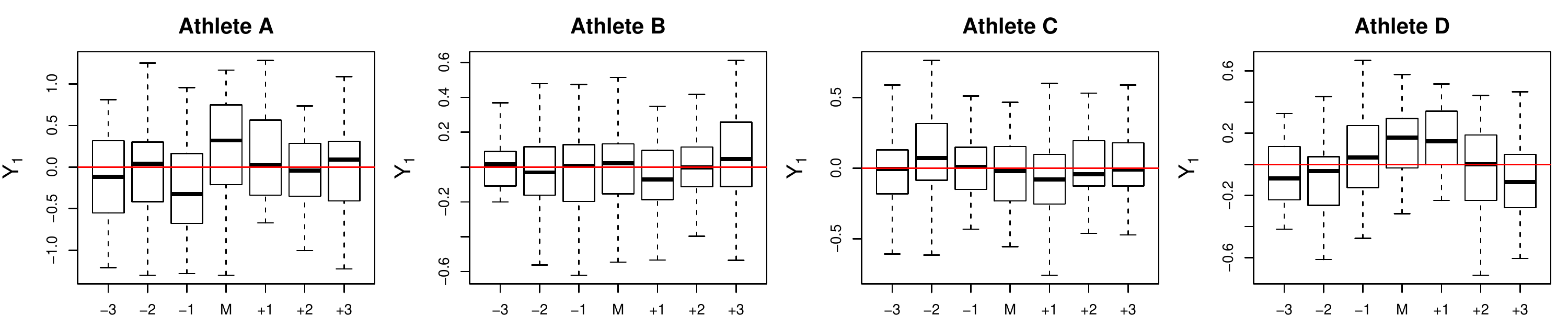}
\caption{Posterior mean estimates of the workload latent factor, $Y_{1it}$, for each individual and match (M), as well as the three days leading up to and following the match. The estimates shown are mean-centered for each match. \label{Fig:LatY1}}
\end{center}
\end{figure}

\begin{figure}
\begin{center}
\includegraphics[scale=.48]{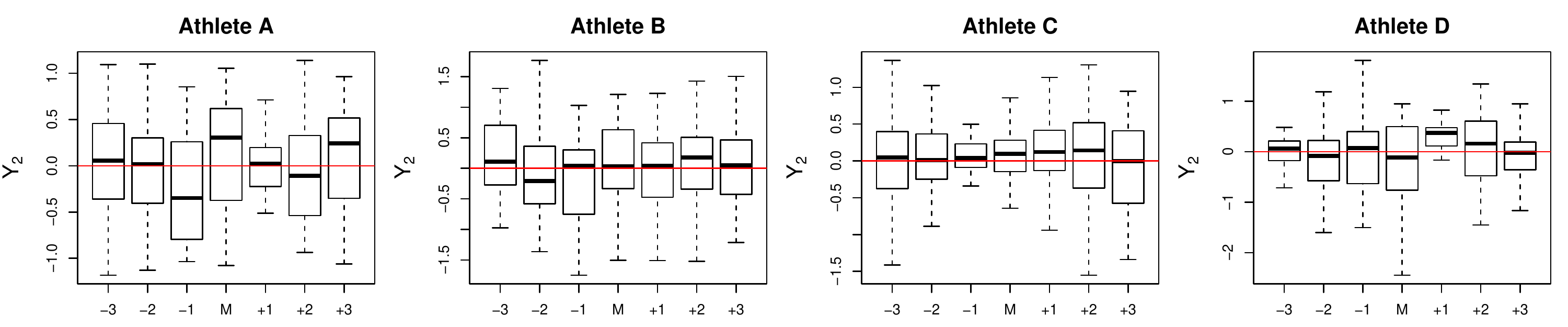}
\caption{Posterior mean estimates of the recovery latent factor, $Y_{2it}$, for each individual and match (M), as well as the three days leading up to and following the match. The estimates shown are mean-centered for each match. \label{Fig:LatY2}}
\end{center}
\end{figure}

\begin{figure}
\begin{center}
\includegraphics[scale=.4]{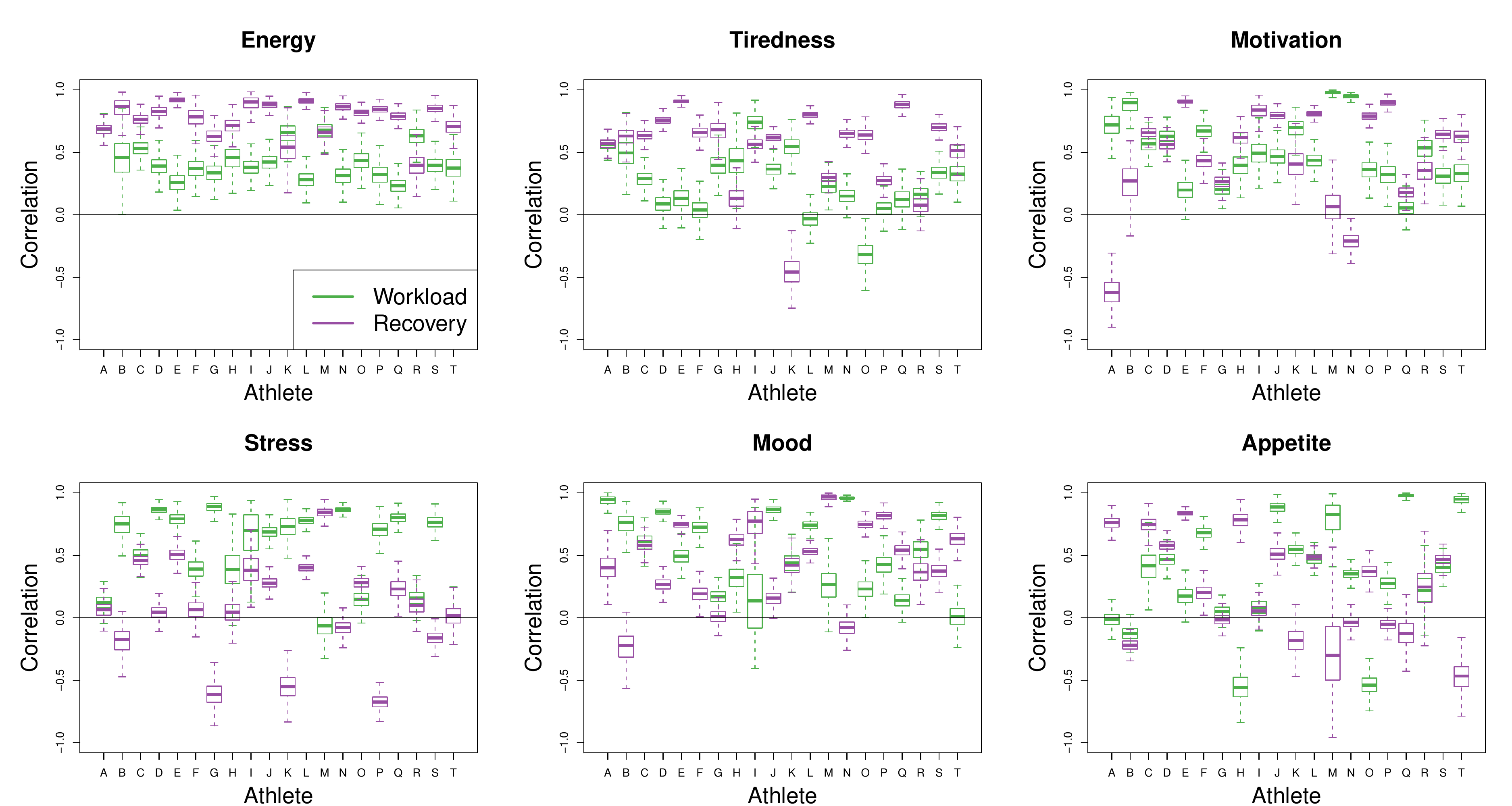}
\caption{Boxplots of the posterior distributions of the correlation between each latent continuous wellness metric, $\widetilde{\mathbf{Z}}_{ij}$, and the latent factors for workload and recovery $\mathbf{Y}_{i1}$ and $\mathbf{Y}_{i2}$, for each athlete $i$ and latent metric variable $j$. \label{Fig:CorrMALL}}
\end{center}
\end{figure}

\begin{figure}
\begin{center}
\includegraphics[scale=.4]{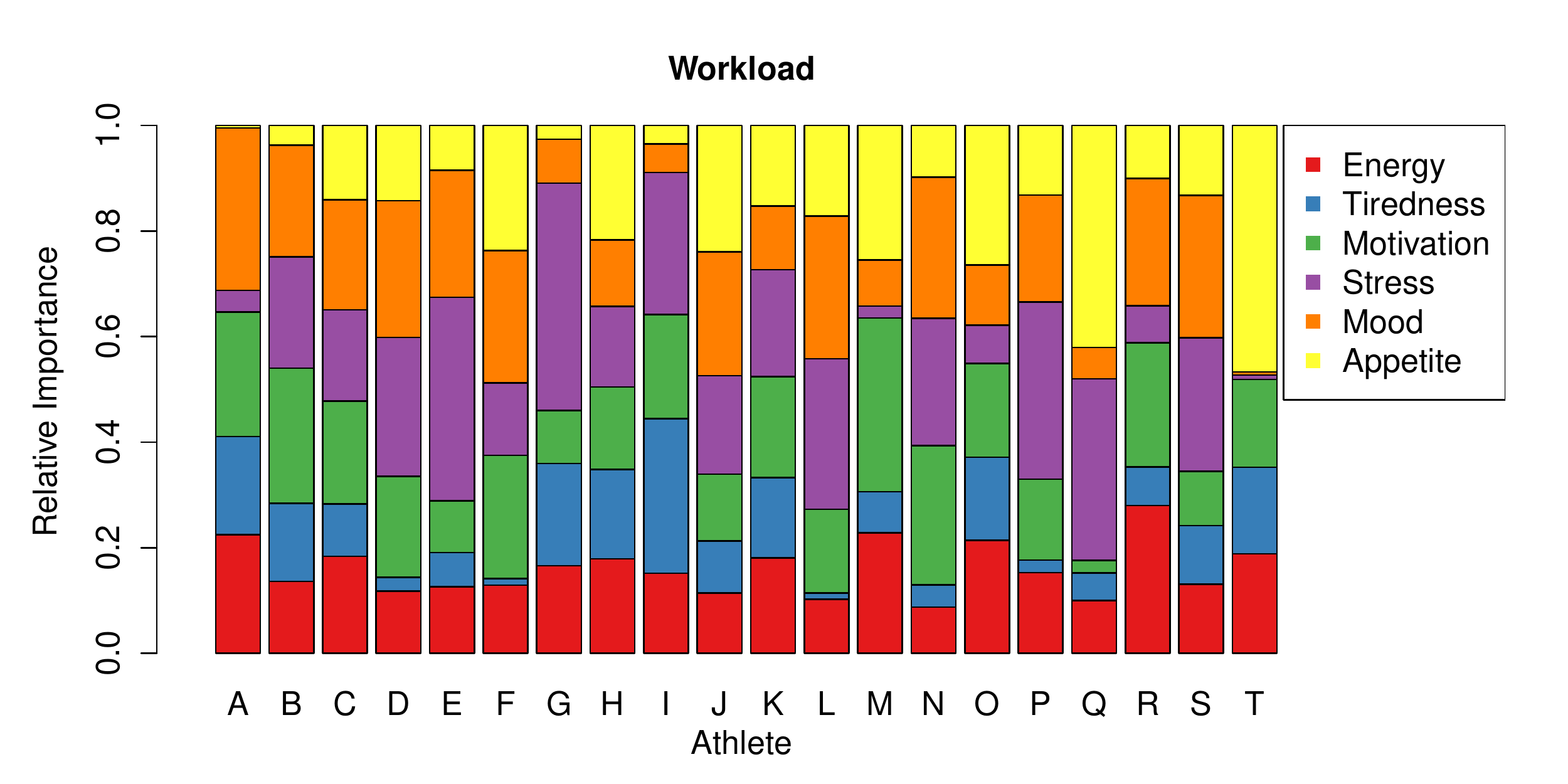}
\includegraphics[scale=.4]{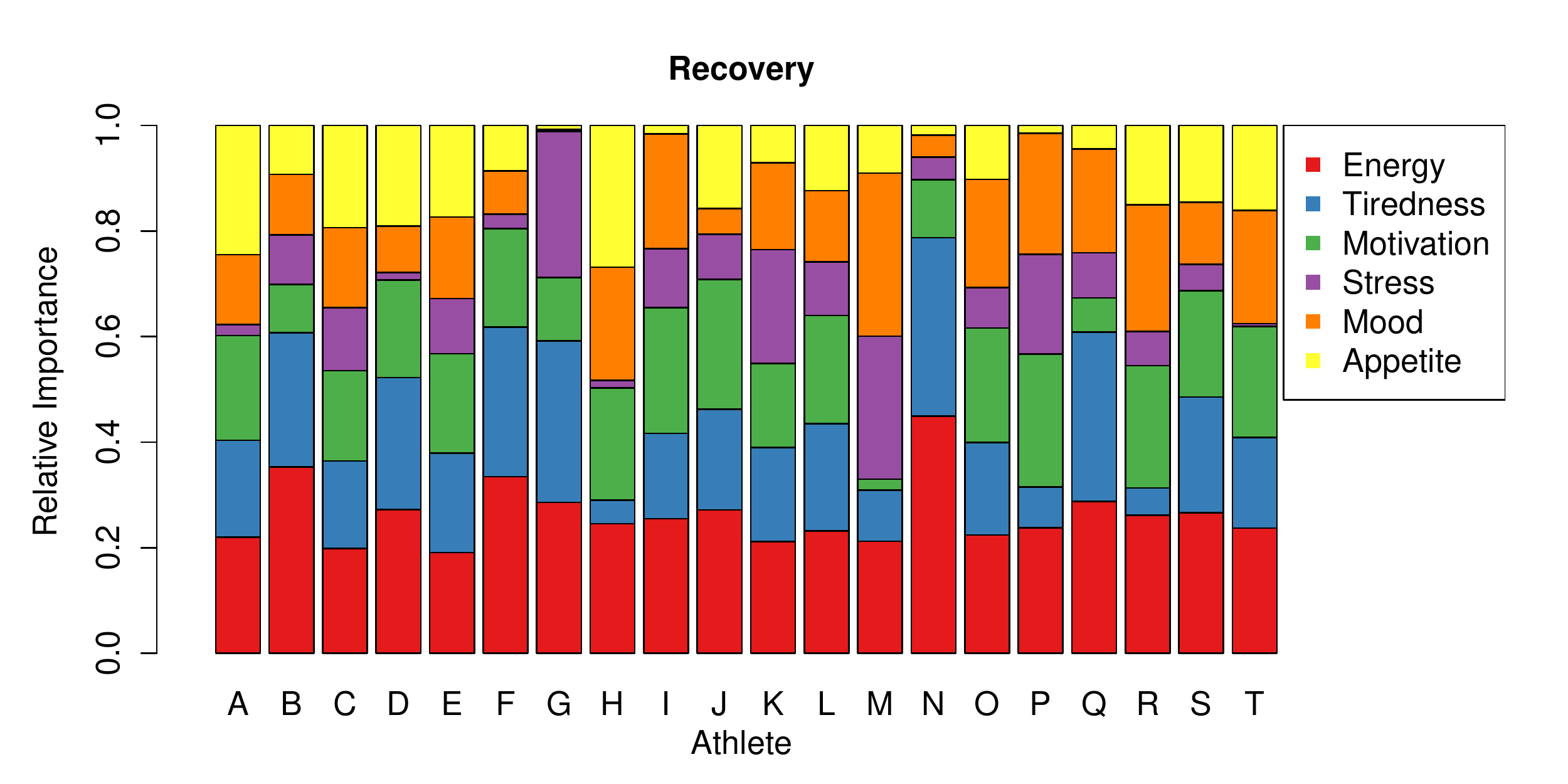}
\caption{Posterior mean estimates of the relative importance statistics, $R_{mj}$, defined in (10) for each athlete and metric for the workload latent factor (top) and recovery latent factor (bottom).\label{Fig:RIMALL}}
\end{center}
\end{figure}

\end{document}